\documentclass[10pt, twocolumn, conference]{IEEEtran}
\IEEEoverridecommandlockouts

\usepackage{subfigure}
\usepackage{paralist}
\usepackage{color}
\usepackage{url}
\usepackage[linesnumbered,ruled,boxed,vlined]{algorithm2e}

\usepackage{amsmath}
\usepackage{amssymb}
\usepackage{graphicx}
\usepackage{bm}
\usepackage{paralist}
\usepackage{subfigure}
\usepackage{times}
\usepackage{balance}
\usepackage{cite}
\usepackage{comment}

\newcommand{\separator}{
  \begin{center}
    \rule{\columnwidth}{0.3mm}
  \end{center}
}

\def\Bl{\Bigl}
\def\Br{\Bigr}

\def\eg{{\it e.g.}}
\def\ie{{\it i.e.}}

\newtheorem{theorem}{Theorem}

\newtheorem{proposition}{Proposition}
\newtheorem{lemma}{Lemma}

\newcommand{\bprob}[1]{\mathbb{P}\Bl[ #1 \Br]}
\newcommand{\prob}[1]{\mathbb{P}[ #1 ]}

\newcommand{\beq}{\begin{eqnarray*}}
\newcommand{\eeq}{\end{eqnarray*}}
\newcommand{\beqn}{\begin{eqnarray}}
\newcommand{\eeqn}{\end{eqnarray}}
\newcommand{\bemn}{\begin{multiline}}
\newcommand{\eemn}{\end{multiline}}

\newcommand{\note}[1]{{\color{red}{[\textit{#1}}]}}

\newcommand\etal{{\em et al.}}

\allowdisplaybreaks
\begin{document}

\title{Rumor Source Detection under Querying with Untruthful Answers %  \thanks{
    %     ACK
    %     %This work was
    % %supported by the National Research Foundation of Korea (NRF) grant
    % %funded by the Korea government (MSIP) (NRF-2013R1A2A2A01067633,
    % %NRF-2013R1A1A3A04007104)
    %     	}
              }

%%\author
%%{Jaeyoung Choi, Sangwoo Moon, Jiin Woo, KyunHwan Son, Jinwoo Shin and Yung Yi\\
%%\IEEEauthorblockA{Department of Electrical Engineering\\
%%KAIST, Republic of Korea\\ Emails: \{jychoi14,mununum,jinwoos,yiyung\}@kaist.ac.kr}
%%\thanks{This work was supported by Institute for Information \& communications Technology Promotion (IITP) grant funded by the Korea government (MSIP) (No.B0717-16-0034,Versatile Network System Architecture for Multi-dimensional Diversity). }
%%}
\author
{Jaeyoung Choi, Sangwoo Moon, Jiin Woo, Kyunghwan Son, Jinwoo Shin and Yung Yi$^\dag$
\thanks{ $\dag$: Department of EE, KAIST,
Republic of Korea. (e-mails: \{jychoi14,mununum,jidoll333,kevinson9473,jinwoos,yiyung\}@kaist.ac.kr }
\thanks{This work was supported in part by the National Research Foundation of Korea (NRF) grant funded by the Korea government (MSIP) (No. 2016R1A2A2A05921755) and Institute for Information and communications
Technology Promotion (IITP) grant funded by the Korea government (MSIP)
(No.B0717-17-0034, Versatile Network System Architecture for Multidimensional
Diversity). 
}
}

\maketitle

\begin{abstract}
  Social networks are the major routes for most individuals to exchange
  their opinions about new products, social trends and political issues
  via their interactions.  It is often of significant importance to
  figure out who initially diffuses the information, \ie, finding a
  rumor source or a trend setter. It is known that
  such a task is highly challenging and the source detection
  probability cannot be beyond $31\%$ for regular trees, if we just
  estimate the source from a given diffusion snapshot.  In practice,
  finding the source often entails the process of querying that asks
  ``Are you the rumor source?'' or ``Who tells you the rumor?'' that
  would increase the chance of detecting the source.  In this paper, we
  consider two kinds of querying: (a) simple \emph{batch} querying and
  (b) \emph{interactive} querying with \emph{direction} under the
  assumption that queriees can be untruthful with some probability.  We
  propose estimation algorithms for those queries, and quantify
  their detection performance and the amount of extra budget due to
  untruthfulness, analytically showing that querying significantly
  improves the detection performance. We perform extensive simulations
  to validate our theoretical findings over synthetic and real-world
  social network topologies.

% As a natural step to increase the detection probability, a source
%   finder may ask questions to those who have already infected.  In
%   this paper, we study the querying scheme such as `whether you are
%   a source or not' or `where you heard a rumor from' with noisy
%   feedback because the feedback answer may have the noisy in
%   practice. To this end, we consider two querying scenarios with a
%   given budget $K$: (a) interactive and (b) non-interactive.  In
%   practice, the answer of the query may has the noise with some
%   probability. We first introduce the querying algorithms for both
%   scenarios, and provide mathematical characterizations in terms of
%   the degree of contributions made by queries, under regular tree
%   topologies. Furthermore, we execute extensive simulations by
%   considering the general graph topologies which show how these two
%   querying schemes performs under the noisy feedbacks, respectively.

\end{abstract}

\section{Introduction}

Information spread is universal in many types of online/offline and
social/physical networks. Examples include the propagation of infectious
diseases, the technology diffusion, the computer virus/spam infection in
the Internet, and tweeting and retweeting of popular topics.  Finding a
``culprit'' of the information spreading is of great significance,
because, for a harmful diffusion, its spreading can be mitigated or even
blocked by vaccinating humans or installing security updates.  Detecting
the rumor source has been regarded as a challenging task unless
sufficient side information is provided. The seminal work by Shah and
Zaman~\cite{shah2010} analytically provides the detection performance of
the MLE (maximum-likelihood estimator) under regular tree topologies,
where the detection probability is upper-bounded by 31\% if the number
of infected nodes goes to infinity and much less
for other practical topologies.  Since then, extensive attentions have
recently been made in various types of network topologies and diffusion
models \cite{shah2012,zhu2013,Luo2013,bubeck2014}, whose major interests
lie in constructing an efficient estimator and providing theoretical
limits on the detection performance.

In practice, the effort of finding the rumor source is made in
conjunction with extra processes with the aim of obtaining more side
information and thus improving the detection performance. In this
paper, we aim at quantifying the impact of querying, where querying
refers to the process of asking some questions. Obviously, it is
expected that such queries improves the detection performance, but
little attention has been made to quantification of the detection
performance in presence of querying. In literature, it has been
studied what happens if multiple snapshot observations are provided
\cite{Zhang2014}, or if a restricted node subset (also called a
suspect set) is given \cite{dong2013} a priori.

In this paper, we study the impact of querying in a highly generalized
setup. Users\footnote{In this paper, the terms ``user" and ``queriee" are used interchangeably. }
may be untruthful with some probability, where two different
types of querying are considered: {\em (a) simple batch querying} and {\em (b)
  interactive querying with direction.} In simple batch querying, for a
given querying budget, a candidate queriee set is first chosen, and the
question of ``Are you the rumor source?'' (referred to as {\em identity question}) can be asked to the queriees
in the set multiple times. Due to limited budget, a source estimation
algorithm should strike a good balance between the size of the candidate
set and the number of questions, depending on the amount of budget, the
degree of untruthfulness, and the underlying graph topology.  In
interactive querying with direction, we start with some initial quieree,
and iteratively ask a series of questions ``Are you the rumor source?''
and ``If not, who spreads the rumor to you?'' (referred to as {\em
  direction question}) to the current queriee,
and determine the next queriee, using the (possibly untruthful) answers
for the second question, where this iterative querying process lasts
until the entire querying budget $K$ is exhausted. A source estimation
algorithm in this query type should smartly consider the tradeoff
between the number of questions and the number of queriees we can ask.

We propose the estimation algorithms for both types of queries and
analyze their detection performances as well as the minimum budget to
satisfy an arbitrary detection performance.  We summarize our main
contributions as follows:
\begin{itemize}
\item[$\circ$] {\bf \em Simple batch querying.}  We first formulate an
  optimization problem that maximizes the detection probability over the
  number of questions to be asked, the
  candidate queiree set, and the estimators for a given diffusion
  snapshot and the answer samples. We discuss its analytical challenges
  and propose an approximate estimation algorithm that selects the
  candidate queriee set based on the hop-distance from the rumor center
  and the MLE for the ``filtered'' candidate nodes by including
  only nodes with many positive answers.  We prove that for a given
  probability $p >1/2$ that users are truthful, $d$-regular tree, and a
  budget $K,$ for any $0< \delta <1,$ if
  $K\geq \frac{c(2/\delta)}{(p-1/2)^{2}\log (\log(2/\delta))}$ with some
  constant $c$ (depending on the degree $d$) then the detection
  probability is at least $1-\delta.$

% we first consider the
%   following two cases: (a) the queried node answers the truth (the fact
%   that she is the source or not) with probability $p >1/2$ and (b) the
%   querier can ask multiple times to the node in the candidate set for
%   improving the quality of the answer. Then, by considering these facts,
%   we design an algorithm which is simple and easily applicable as well
%   as high quality for detecting the source. We analyze the detection
%   probability for $d$-regular tree for given $K$ and $p$ and we derive
%   that if $K\geq \frac{c(2/\delta)}{p^{2}\log (\log(2/\delta))}$ with
%   some constant $c$ depends on the degree $d$ then the detection
%   probability is at least $1-\delta$ for arbitrary small $\delta>0$ for
%   this querying scenario.

%  We consider two querying types: (i) {\em interactive} and {\em
%  non-interactive,} where some query budget $K$ is given.
%  Non-interactive querying corresponds to a batch type of querying
%  without any interactive question-and-answer mechanism. One just
%  chooses $K$ candidate nodes at once and perform a single query to
%  each of them, asking `Are you a rumor source?'. Another type is
%  interactive querying that asks $K$ candidate nodes sequentially,
%  where the questions is `who spreads a rumor to you?'.  Note that
%  one can obtain additional information in each query, which can be
%  utilized to choose the next queriee.

\smallskip
\item[$\circ$] {\bf \em Interactive querying with directions.}  In this
  querying type, we also consider an optimization problem that maximizes
  the detection probability and discuss the technical challenges in
  solving it, where we assume that users are only untruthful for the
  direction question for simplicity. We propose an estimation algorithm
  that starts with a ``rumor center" as the initial queriee, and apply a
  kind of majority rule in determining the next queriee, \ie, selecting
  the node with highest vote for multiple direction questions. We
  analyze this simple, yet powerful estimation algorithm and
  characterize its detection probability for given parameters. From
  this, we establish the minimum budget for any given detection
  probability: for the probability $q > 1/d$ that users are truthful for
  a direction question, $d$-regular tree, and a given budget $K,$ for
  any $0< \delta <1,$ if
  $K\geq \frac{c\log(7/\delta)}{(q-1/d)^3\log (\log (7/\delta))}$ with some
  constant $c$ (which depends on the degree $d$), then the detection
  probability is at least $1-\delta.$ This result quantifies the power
  of the direction question in addition to the identity question, reducing
  the required budget to satisfy the detection probability $1-\delta$ by
  a logarithmic factor with respect to the scaling of $1/\delta$ for
  small $\delta >0.$

% For
%   the interactive querying with direction, a querier chooses the
%   queriee iteratively, where who will be the $k$-th queriee may be
%   determined by the answer of $(k-1)$-th queriee. This setting assumes
%   that the querier can ask an additional direction query \ie, ``From
%   whom you have heard the rumor?'' if the queried node is not the
%   source.  In this querying scheme, we also consider (a) the untruthful
%   answer of the queried node for the direction question with
%   probability $q>1/d$ for $d$-regular tree and (b) the querier
%   can ask multiple queries for this direction questions for a given $K$. Based on
%   these facts, we also derive the lower bound of detection probability
%   for given $K$ and $q$ for $d$-regular tree and we obtain that if
%   $K\geq \frac{c\log(7/\delta)}{q^3\log (\log (7/\delta))}$ with some
%   constant $c$ depends on the degree $d$ then the detection
%   probability is at least $1-\delta$ for arbitrary small
%   $\delta>0$. This result shows that if the querier has the direction
%   information of the source, then the order of number of queries
%   deducted by logarithm fact.  Our theoretical results give guide
%   lines for the required number of querying budget when the target
%   detection probability is given under untruthful feedback.

\smallskip
\item[$\circ$] {\bf \em Evaluation over synthetic and real-world
    graphs.} Our analytical results above provide useful guidelines on
  how much budget is required to guarantee a given detection performance
  for different querying types when users are untruthful. We validate
  our findings via extensive simulations over popular random graphs
  (\emph{Erd\"{o}s-R\'{e}nyi}  and scale-free graphs) and a real-world
  Facebook network. As an example, in Facebook network,
the interactive querying requires about 200 queries to achieve almost
one detection probability when $q>0.5$ because
the tracking by the direction is efficient due to the small diameter of the
network.
\end{itemize}
 \vspace{-0.4cm}

\bigskip
\noindent{\bf \em Related work.}
%\smallskip
% source detection: point estimator
The research on rumor source detection has recently received
significant attentions. % In this section we briefly review some related
% works that tried to solve the rumor source detection problem, in
% various settings and environments.
The first theoretical approach was done by Shah and Zaman
\cite{shah2010,shah2012,shah2010tit} and they introduced the metric
called {\em rumor centrality}, which is a simple topology-dependent
metric.  They proved that the rumor centrality describes the likelihood
function when the underlying network is a regular tree and the diffusion follows
the SI (Susceptible-Infected) model, which is
extended to a random graph network in \cite{shah2011}.  Zhu and Ying
\cite{zhu2013} solved the rumor source detection problem under the SIR
(Susceptible-Infected-Removed) model and took a sample path approach to
solve the problem, where a notion of {\em Jordan center} was introduced,
being extended to the case of sparse observations \cite{zhu2014}. The
authors of \cite{bubeck2014}, \cite{fuchs2015} and \cite{frieze2014}
studied the problem of estimating the source for random growing trees,
where unlike aforementioned papers, they did not assume an underlying
network structure. The authors in \cite{farajtabar2015} inferred the
historical diffusion traces and identifies the diffusion source from
partially observed cascades, and similarly in \cite{pinto2012}, partial
diffusion information is utilized. Recently, there has been some
approaches for the general graphs in \cite{Kai2016, Chang2015} to find
the information source of epidemic. All the detection mechanisms so far
correspond to point estimators, whose detection performance tends to be
low.  There was several attempts to boost up the detection probability.
Wang \etal \cite{Zhang2014} showed that observing multiple different
epidemic instances can significantly increase the the detection
probability. Dong \etal \cite{dong2013} assumed that there exist a
restricted set of source candidates, where they showed the increased
detection probability based on the MAPE (maximum a posterior estimator).
Choi \etal \cite{Choi2016,Choi2016w} showed that the anti-rumor spreading under
some distance distribution of rumor and anti-rumor sources helps finding
the rumor source by using the MAPE.
The authors in \cite{bubeck2014, Khim14,Moon2016} introduced the notion of {\em
  set estimation} and provide the analytical results on the detection
performance. These are close to our work, where querying is considered
in detecting the rumor source. However, our work is done in a much more
generalized and practical setup in the sense that we consider the case
when users may be untruthful, and also two types of practical querying
scenarios are studied. %In fact, some of the results in \cite{Khim14}
\section{Model and Preliminaries}
\label{sec:model}

\subsection{Model}
\vspace{-0.06cm}
\noindent{\bf \em Rumor diffusion.} We describe a rumor spreading
model which is commonly adopted in other related work, \eg,
\cite{shah2010,shah2012,zhu2013}.  We consider an undirected graph
$G=(V,E),$ where $V$ is a countably infinite set of nodes and $E$ is
the set of edges of the form $(i,j)$ for $i, j\in V$.  Each node
represents an individual in human social networks or a computer host
in the Internet, and each edge corresponds to a social relationship
between two individuals or a physical connection between two Internet
hosts.  We assume a countably infinite set of nodes for avoiding the
boundary effect. As a rumor spreading model, we consider a SI model,
where each node is in either of two states: {\em susceptible} or {\em
  infected}. All nodes are initialized to be susceptible except the
rumor source, and once a node $i$ has a rumor, it is able to spread
the rumor to another node $j$ if and only if there is an edge between
them.  Let a random variable $\tau_{ij}$ be the time it takes for node
$j$ to receive the rumor from node $i$ if $i$ has the rumor. We assume
the $\tau_{ij}$ are exponentially distributed with rate $\lambda>0$
independently of everything else. Without loss of generality, we
assume that $\lambda=1$. We denote $v_1 \in V$ by the rumor source, which
acts as a node that initiates diffusion and denote $V_N \subset V$ by $N$ infected
nodes under the observed snapshot $G_N\subset G$. In this paper, we consider
the case when $G$ is a regular tree
and our interest is when $N$ is large, as done in many prior work
\cite{shah2010,shah2012,Khim14,dong2013,Zhang2014}.

\begin{figure}
\begin{center}
\subfigure[A simple batch querying.]{\includegraphics[width=0.24\textwidth]{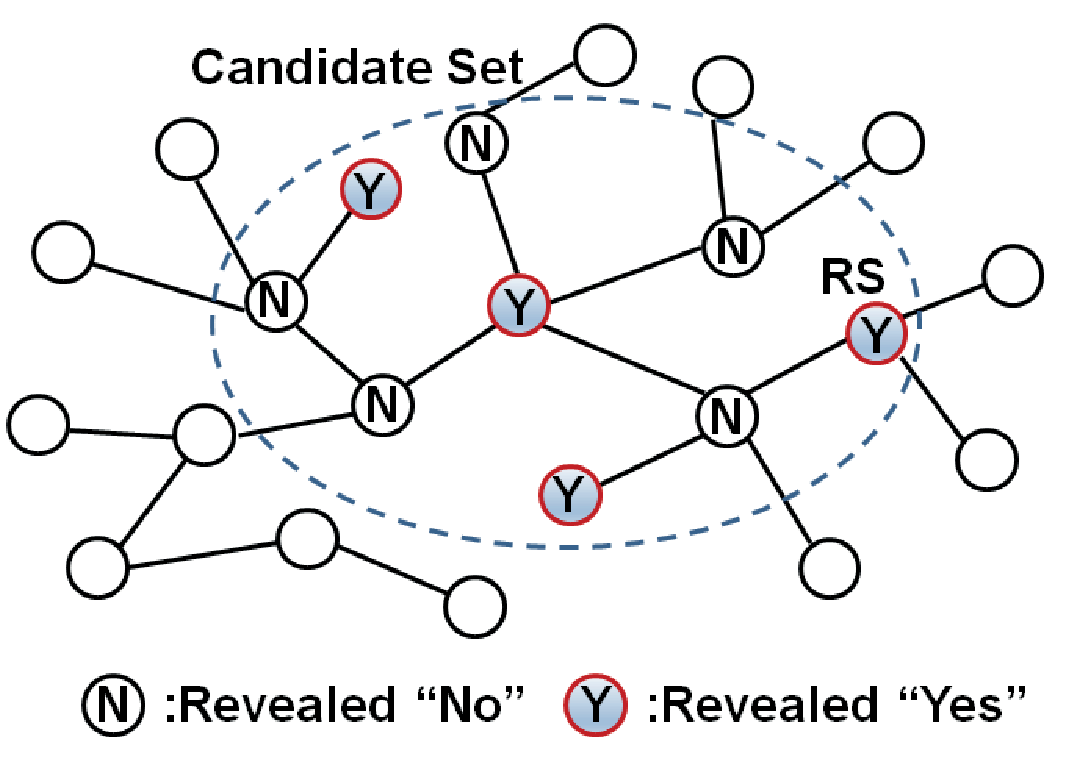}\label{fig:Noninter}}
\hspace{-0.2cm}
\subfigure[Interactive querying with direction.]{\includegraphics[width=0.24\textwidth]{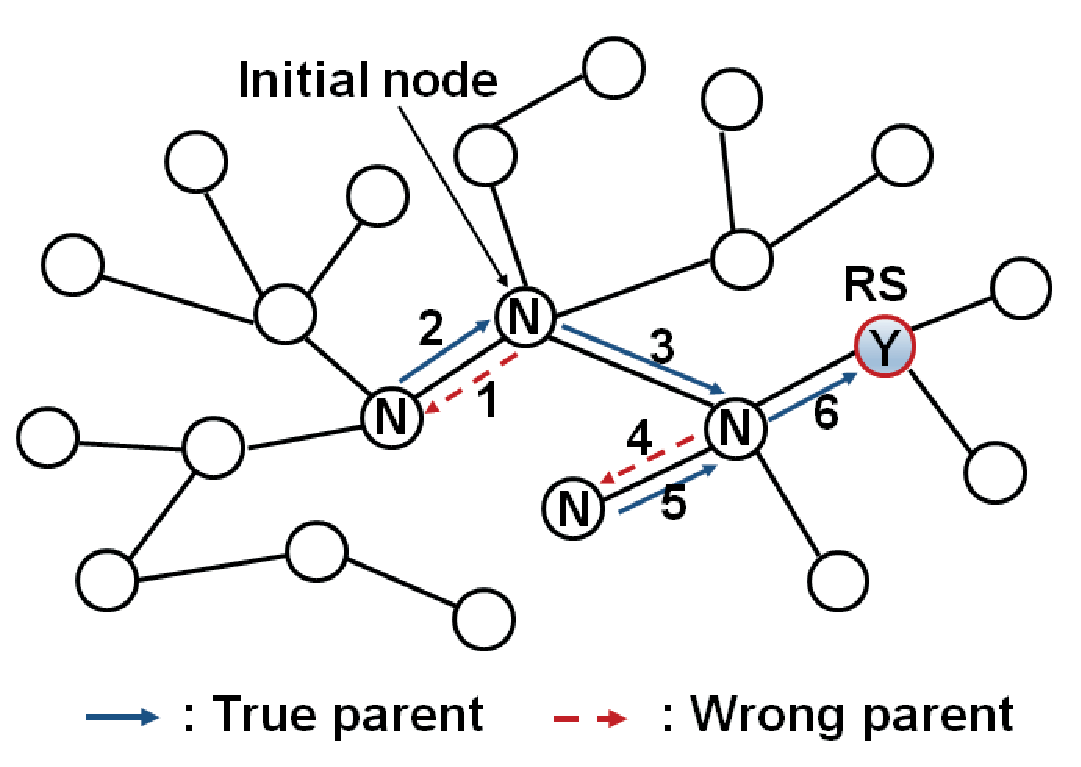}\label{fig:Inter}}
\hspace{0cm}
 \end{center}
\caption{Examples of two querying types with untruthful
  answers. In (a), the querier selects a candidate set (a dotted
  circle) and asks just one identity question (\ie, $r=1)$ in a batch. In (b),
  starting from the initial node, the querier first asks ``Are you the
  rumor source?'', and if no, she further asks ``Who spreads the rumor to
  you?'', where the querier interactively tracks the true source, but the
  queriees may be untruthful (RS: Rumor Source). }
\label{fig:querying}
% \vspace{-0.5cm}
\end{figure}

%\begin{figure}
%\centerline{\includegraphics[width=0.3\textwidth]{figs/Noninter}} \caption{Explanation
%of $T^v_u$. $T^1_2$ is 3 because there are 3 nodes in the subtree
%rooted at node 2 considered node 1 is the overall root of the
%tree.}  \label{fig:subtree} \end{figure}

% \subsection{Querying Scenario}

\smallskip
\noindent{\bf \em Querying with untruthful answers.}  A detector is
allowed to query the nodes, where querying refers to a process of
asking some questions (which will be shortly clarified depending on querying scenarios).
The detector is given some querying budget
$K,$ where we assume that one budget is used to ask one question.  In
this paper, we consider two types of queries (see
Fig.~\ref{fig:querying}), where a queriee may be untruthful, as
modeled in what follows:

% of a particular node whether she is a rumor source or not, where we
% assume that the queriee is truthful in answering questions. It is
% clear that more queries would help to detect the rumor source and
% the main goal in this paper is to quantify its effect. We consider
% the following two querying scenarios.

%  with a given budget $K,$ where a unit budget is used to ask a unit
% of questions.

% % Depending on how querying is performed, we consider the following two
% % querying scenarios:

\smallskip
\begin{compactenum}[$\circ$]
\item {\bf \em Simple batch querying.}  This query is parameterized by
  $r$, where a querier first chooses $K/r$  ($K$ is a multiple of $r$ for expositional convenience)
  candidate nodes to each of which an {\em identity question} of ``Are
  you the rumor source?'' is asked $r$ times. We call $r$ the
  {\em repetition count} throughout this paper.
  Each queriee $v$ is
  truthful in answering each question only with probability $p_v,$ \ie,
  even if she is the rumor source, she lies with probability $1-p_v.$
  We assume that the answers are independent over the $r$ queries across
  the queriees. We assume the homogeneous case when $p_v = p$ for all
  $v \in V_N,$ and $p > 1/2,$ \footnote{In the case $p<1/2$, one can flip the sign of the
  final estimation to achieve the same guarantee.} \ie, all users are biased with truthful
  answers. For example, Fig~\ref{fig:Noninter} shows a
  candidate set of nodes inside a dotted circle, where with $r=1$ we
  have four ``yes'' nodes and five ``no'' nodes.

\smallskip
\item {\bf \em Interactive querying with direction.}  In this querying
  type, there are two questions, where one is the identity question, as
  in the simple batch querying and another is the {\em direction
    question} of ``Who spreads the rumor to you?.''  This query is also
  parameterized by $r,$ where the querying process occurs in the
  following interactive manner: A querier first chooses an initial node
  to ask the identity question, and further asks the direction question
  $r$ times (\ie, repetition count), if the queriee answers that she is
  not the rumor source.  The querying process stops when the queriee
  says that she is the rumor source or the entire budget $K$ is
  exhausted. A querier determines the next queriee from $r$ direction
  questions, where each queriee $v$ is truthful for the direction
  question only with probability $q_v.$\footnote{We model untruthfulness
    only in the direction question, because we aim at purely focusing on
    the impact of untruthfulness in the interactive search of the true
    rumor source, which seems to be the critical component in this type
    of querying.} In other words, she lies for the direction question
  with probability $1-q_v$ and designates a node uniformly at random out
  of all neighbors except for the node who truly spreads the rumor to
  $v$ as a bogus ``parent''. Then, the querier chooses one of her
  neighbors as the parent of $v$ and repeats the same procedure.  As in
  the simple batch querying, we assume the homogeneity in truthfulness
  that $q=q_v$ for all $v \in V_N,$ and $q > 1/d,$ \ie, users' bias for
  truthful answers\footnote{In $d$-regular trees, when $q=1/d$, the
    answer for the direction question turns out to be uniformly
    random.}.  Fig~\ref{fig:Inter} shows an example scenario that
  starting from the initial node, a sequence of nodes answer the
  interactive queries truthfully or untruthfully.

% ----------------------- If the node reveals itself is not the rumor
%   source then ask \emph{additional} queries as ``From whom you have
%   heard the rumor?'' which is an {\em direction question}, $r$
%   times.  Each queriee $v$ is truthful in answering for the identity
%   question and is truthful in answering for the direction question
%   with probability $q_v,$ i.e., she lies with probability $1-q_v$
%   for saying its parent node \footnote{For simplifying the analysis,
%   we assume that the queried node answer the truth for the identity
%   question in the interactive querying scenario.}.  We also assume
%   that the answers of direction questions are independent across the
%   $r$ queries. In this setting, a querier chooses the queriee
%   iteratively, where who will be the $k$-th queriee may be
%   determined by the answer of $(k-1)$-th queriee.

\end{compactenum}

% In practice, the queried node may not give the correct answer for
% the questions so that we need to consider the scenarios that the
% feedback of all queries are not true \ie some of them has a noise
% with their answer.  In this paper, we consider the following
% feedback scenario.  \smallskip \begin{compactitem}[$\circ$] \item
% {\bf Feedback scenario.} When an infected node $v$ is selected by a
% querier, the node reveals the correct answer with probability
% $p_v>0$ for the first question (Are you the rumor source ?) in the
% non-interactive scenario. For the interactive case, the node reveals
% the correct answer for the first question but reveals the correct
% answer with probability $q_v >0$ for the second question (From whom
% you have heard the rumor ?), respectively\footnote{In our work, we
% do not consider the noisy feedback for the first question in the
% interactive scenario for simple analysis.}.  Thus, the rumor source
% detection without queries in most prior works corresponds to the
% case when $K=1$ and $p_v=1$ for all $v\in V_N$.

\smallskip
\noindent{\bf \em Goal.} Our goal is to propose efficient estimation
algorithms that are practically implementable, for both types of queries
with users' untruthfulness. Especially, we aim at theoretically
quantifying the detection performance of our proposed algorithms by
providing the lower bound of the required budget $K$ to satisfy any
arbitrary detection probability, \ie, the sufficient budget $K$ for the
target detection quality.
\vspace{-0.1cm}

% The goal of this paper is to quantitatively
% understanding how much querying helps in improving the detection
% probability for both querying scenarios with unfaithful answers.
% To do this, designing simple and smart algorithms is a challenging
% issue to guarantee the high source detection performance. In the
% following sections, we introduce the source estimation algorithm
% for both querying scenarios to resolve these issues. Before seeing
% this, we remind a preliminary result which is used in our algorithms.
\subsection{Preliminaries: Rumor Centrality}
\vspace{-0.06cm}
\label{sec:rumor_center}
\vspace{-0.06cm}
As a preliminary, we explain the notion of {\em rumor centrality,} which is a
graph-theoretic score metric and is originally used in detecting the
rumor source in absence of querying and users' untruthfulness, see
\cite{shah2010}. This notion is also importantly used in our framework
as a sub-component of the algorithms for both simple batch querying and
interactive querying with direction.
% In this section, we provide the known results for rumor source
% detection in absence of querying, \ie, $K=1$.  One of the most
% popular estimators is based on maximum likelihood (ML).
In regular tree graphs, Shah and Zaman \cite{shah2010} showed that the
source chosen by the MLE becomes the node
with highest rumor centrality. Formally, the estimator chooses $v_{RC}$ as
the rumor source defined as
\begin{eqnarray}\label{eqn:ML}
       \vspace{-0.4cm}
v_{RC} &=& \arg\max_{v \in V_N} \mathbb{P}(G_N | v=v_1)\cr
       &=&\arg\max_{v \in V_N} R(v, G_N),
              \vspace{-0.4cm}
\end{eqnarray}
where $v_{RC}$ is called \emph{rumor center} and $R(v, G_N)$ is the
rumor centrality of a node $v$ in $V_N$. The rumor centrality of a
particular node is calculated only by understanding the graphical
structure of the rumor spreading snapshot, \ie, $R(v, G_N) =N!
\prod_{u \in V_N} (1/T^v_u)$ where $T^v_u$ denotes the number of nodes
in the subtree rooted at node $u$, assuming $v$ is the root of tree
$G_N$ (see \cite{shah2010} for details).
% \footnote{Refer \cite{shah2010} to see the details about the rumor
% centrality in the $d$-regular tree.}

%%% Local Variables:
%%% mode: latex
%%% TeX-master: "main"
%%% End:

% \input{noninteractive}

\section{Detection Using Simple Batch Queries}
\label{sec:noninteractive}
\vspace{-0.1cm}
\subsection{Algorithm based on MLE and Rumor Centrality}
\label{sec:simple_mle}
\vspace{-0.1cm}
A source estimation algorithm for simple batch queries consists of the
following steps: We first need to appropriately choose the repetition
count $r$ and the candidate set $C_r \subset V_N$ of size $K/r$ and ask the queries
to the nodes in $C_r.$ Then, we will be given a sample of the
answers, which we denote by a vector
$A_r := A_{r}(p)=[x_1, x_2, \ldots, x_{K/r}]$ with $0\leq x_i \leq r$
representing the number of ``yes'' answers of the $i$-th node of $C_r.$
Then, it is natural to consider an algorithm based on MLE, to maximize
the detection probability, that solves the following optimization:
\vspace{-0.3cm}
\separator
\vspace{-0.3cm}
\begin{align}\label{eqn:OPT}
% \displaystyle
\text{\bf OPT-S:} &  \quad \max_{1\leq r \leq K} \max_{C_r}
  \max_{v \in C_r} \bprob{G_N,A_{r}| v=v_1},
\end{align}
where the inner-most max corresponds to the MLE given the
diffusion snapshot $G_N$ and the query answer sample $A_r.$
\vspace{-0.3cm}
\separator

\smallskip
\noindent{\bf \em Challenges.} We now explain the technical challenges
in solving {\bf OPT-S}. To that end, let us
consider the following sub-optimization in {\bf OPT-S} for a fixed $ 1\le
r \le K$:
\begin{align}\label{eqn:subOPT}
% \displaystyle
\text{\bf SUB-OPT-S:} &  \quad \max_{C_r}
  \max_{v \in C_r} \bprob{G_N,A_{r} | v=v_1}.
\end{align}
Then, the following proposition provides the solution of {\bf SUB-OPT}
whose proof is provided in our technical report \cite{Jae16}.

\smallskip
\begin{proposition}\label{pro:subOPT}
  Construct $C_r^*$ by including the $K/r$ nodes in the decreasing order
  of their rumor centralities. Then, $C_r^*$ is the solution of {\bf SUB-OPT-S}.
\end{proposition}

\smallskip
Despite our knowledge of the solution of {\bf SUB-OPT-S},
solving {\bf OPT-S} requires an analytical form of the objective value of
{\bf SUB-OPT-S} for $C_r^*$ to find the optimal repetition count, say
$r^*.$ However, analytically computing the detection probability for
a given general snapshot is highly challenging due to the following
reasons. We first note that
\begin{align}
  \label{eq:trade}
&\max_{v \in C_r} \bprob{G_N,A_{r} | v=v_1} \cr
& =  \underbrace{\prob{v_1 \in  C_r^*}}_{(a)} \times \underbrace{\max_{v \in C_r^*} \bprob{G_N,A_{r}| v=v_1, v_1 \in C_r^*}}_{(b)}.
\end{align}
First, the term $(a)$ is difficult to analyze, because only the rumor
center allows graphical and thus analytical characterization as
discussed in Section~\ref{sec:rumor_center}, but other
nodes with high rumor centrality is difficult to handle due to the
randomness of the diffusion snapshot.
Second, in $(b)$, we observe
that using the independence between $G_N$ and $A_r,$ by letting the
event ${\cal A}(v) = \{v=v_1, v_1 \in C_r^*\},$
\begin{align}
\label{eq:2}
\hat{v} &= \arg \max_{v \in C_r^*} \bprob{G_N,A_{r} \mid  {\cal A}(v)} \cr
&= \arg \max_{v \in C_r^*} \bprob{A_r
  \mid {\cal A}(v)} \times \bprob{G_N\mid {\cal A}(v)}.
\end{align}
Then, the node $\hat{v}$ maximizing $(b)$ is the node $v$ that has the
maximum {\em weighted} rumor centrality where the weight is
$\prob{A_r| {\cal A}(v)}$. As opposed to the case of characterizing the
rumor center in the non-weighted setup \cite{shah2010}, analytically
obtaining or graphically characterizing $\hat{v}$ in this weighted setup
is also hard due to the randomness of the answer for querying,
thus resulting in the challenge of computing $r$ that
maximizes the detection probability in {\bf OPT-S}.

% However, obtaining the exact detection probability
% by considering this weight is quite complex because we need to all possible
% sample data after querying.

%\begin{align}
%  \label{eq:2}
%&  \bprob{v=v_1|G_N,A_{r}, v_1 \in C_r}\cr
%&=  \bprob{v=v_1|G_N, v_1 \in C_r} \times \bprob{v=v_1|A_{r}, v_1 \in
%  C_r},
%\end{align}

% \begin{align}
%   \label{eqn:MAP}
%   \hat{v}_{\tt ml}&=\arg\max_{v \in G_N}\mathbb{P}(G_N,
%                     K^{r}_{p}|v=v_1)\\ &\stackrel{(a)}{=}\arg\max_{v \in
%                                          G_N}\mathbb{P}(G_N|v=v_1)\mathbb{P}(K^{r}_{p}|v=v_1),\\
% \end{align}
% where $(a)$ is from the fact that $G_N$ and $K^{r}_{p}$ are independent.

\begin{comment}
First, it is not easy to find the optimal set $V^r$ for a given budget
$K$ because we do not know what the best strategy to select the set
which maximizes the probability of containing the source. Second, for
calculating the MLE, the first term in \eqref{eqn:MAP} can be obtained
by $R(v,G_N)$ from \eqref{eqn:ML} easily but, the probability
$\mathbb{P}(K^{r}_{p}|v=v_1)$ is quite complex because we have to
consider the all possible cases as varying $r$ which output the data of
each nodes in $V^r$. To handle these issues, we first provide an optimal
solution in \eqref{eqn:OPT} for a given $r$. To see this, consider the
following proposition.
\end{comment}

One can numerically solve {\bf OPT-S}, which, however, needs to generate
a lot of $A_r$ samples (one sample requires a vector of answers for $K$
questions).  Motivated by this, we propose an algorithm producing an
approximate solution of {\bf OPT-S}. The key of our approximate
algorithm is to choose $C_r$ that allows us to analytically compute the
detection probability for a given $r$ so as to compute a good $r$
easily, yet its performance is close to that of {\bf OPT-S}, as
numerically validated in Section~\ref{sec:numerical}.

\subsection{Algorithm based on Hop Distance and Majority Rule}

We now propose an algorithm that overcomes the afore-mentioned
challenges in $(a)$ and $(b)$ of \eqref{eq:trade}.  The key idea is that
for $(a)$ we adopt a {\em hop-distance based selection} of the candidate
set $C_r$ and for $(b)$ we simply apply a {\em majority-based rule}.

We first formally describe our algorithm in {\bf SB-Q}$(r)$
parameterized by a repetition counter $r$, and explain how it operates,
followed by presenting the rationale behind it: we first calculate the
rumor centrality of all nodes in $G_N$ (Line 2), where $s$ is set to be
the rumor center. Then, using the parameter $r,$ we % appropriately choose the repetition count as
% in Line 3, and
construct the candidate set $C_r$ by the nodes within the
hop-distance $l$ given in Line 3 from the rumor center $s$ using the relation
that $K\geq r|C_r|=r\frac{d(d-1)^{l}-2}{d-2}$ (Line 3). Next, for each
node $v$ in $C_r,$ we ask the identity question $r$ times, and count the
number of ``yes''es in $\mu(v)$ (Lines 4-5). Using this, we filter out
the candidate set $C_r$ and construct $\hat{V}$ by including the nodes
with $\mu(v)/r \geq 1/2$ (majority rule), \ie, the nodes with higher
chance to be the rumor source.  When $K$ is not a multiple of $r,$ we
handle the remaining $K-r |C_r|$ nodes as in Lines 6-8. Finally we
choose the node in $\hat{V}$ with highest rumor centrality, where if
$\hat{V} = \emptyset,$ we simply do the same task for $C_r.$

\begin{algorithm}[t!]
 \caption{{\bf SB-Q}$(r)$, $r$: Repetition Count}
\label{alg:noninteractive}
 \KwIn{Diffusion snapshot $G_N$, budget $K$, degree $d$, and truthfulness probability
   $p>1/2$} \KwOut{Estimator $\hat{v}$}
% \vspace{-0.5cm}\separator
$C_r = \hat{V} = \emptyset$\;
Calculate the rumor centrality $R(v, G_N)$ for all $v \in G_N$ as in
\cite{shah2010} and let $ s \leftarrow \arg\max_{v \in V_N} R(v, G_N)$\;
% Set the repetition count $ r \leftarrow \left \lfloor 1+\frac{(1-p)\log K}{2e\log(d-1)}\right\rfloor$ \;
% $ s \leftarrow \arg\max_{v \in V_N} R(v, G_N)$\;
Construct a candidate
set $C_r$ by including each node $v$ that satisfies $d(v,s)\leq l
,$ where $l=\frac{\log\left(\frac{K(d-2)}{rd}+2\right)}{\log(d-1)}$ and
$d(v,s)$ is the hop distance between nodes $v$ and $s$\;
 % Compute the maximum distance $l$ from \eqref{eqn:l} and assign a
 % set $V_l$ which consists of satisfying $d(v,s)\leq l$ for all nodes
 % $v \in V_N$\;
 % l=\frac{\log\left(\frac{K(d-2)}{rd}+2\right)}{\log(d-1)}.

% $\hat{V} \leftarrow \emptyset$ \;
\For{each $v \in C_r$}{
% $\mu(v)\leftarrow 0$\;
Count the number of ``yes'' (\ie, I am the rumor source) for the
identity question, stored at $\mu(v),$ and if $\mu(v)/r \geq 1/2$ then
include $v$ in $\hat{V}$\; }
% \While{$r\geq 1$}{ $v$ selects a random variable $X\sim U(0,1)$\;
%    \If{$X\leq v(p)$}{ $\mu(v)\leftarrow \mu (v) +1$\;} }

%   \If{$\mu(v)/r \geq 1/2$}{ $\hat{V} \leftarrow \hat{V} \cup
%    \{v\}$\;} }
\While{$K-r|C_r|\geq r$}{ Select a node $v$ satisfying $d(v,s)= l+1$
  uniformly at random\; Do the same procedure in Lines 5\; }

 \If{$\hat{V}=\emptyset$}{ $ \hat{v} \leftarrow \arg\max_{v\in C_r}
   R(v, G_N)$\; } \Else{ $ \hat{v} \leftarrow \arg\max_{v\in\hat{V}}
   R(v, G_N)$\; }
% Return $\hat{v}=s$\;
\end{algorithm}

\smallskip
\noindent{\bf \em Rationale.} We now provide the rationale of {\bf SB-Q}$(r)$ from the perspective of how we handle the
analytical challenges in \eqref{eq:trade}  so as to solve
{\bf OPT-S} in an approximate manner.

\smallskip
\begin{itemize}
\item[$\circ$] {\em Hop Distance based $C_r$ selection:} Selecting
  $C_r$ based on the distance from the rumor center, rather than based
  on the sorted rumor centrality permits us to have the closed form of
  $\prob{v_1 \in C_r}.$ However, it is not
  difficult to obtain the lower bound of this probability when we consider
  the hop distance based $C_r$ by using some
  preliminary results in \cite{shah2012, Khim14}. Furthermore, the authors
  in \cite{Khim14} shows the probability of distance between rumor source and center
  decays exponential with respect to the distance, \ie, the source is
  nearby the rumor center with high probability. Hence, it is a good
  approximation to the centrality based $C_r$ with analytical guarantees.

\item[$\circ$] {\em Construction of the filtered set $\hat{V}$ from
    querying:} Consider the answer sample of node $v$ for $r$ questions,
  $x_v ~(1\leq x_v \leq r),$ where one can easily check that for
  $x_v \geq r/2$ then the weight $\prob{A_r| {\cal A}(v)}$ becomes
  larger than that for $x_v < r/2$ due to $p>1/2.$ We use an
  approximated version of the weight from the answer samples by setting
  $\prob{A_r|{\cal A}(v)}=1$ if $x_v \geq r/2,$ and
  $\prob{A_r|{\cal A}(v)}=0$ if $x_v < r/2.$
\end{itemize}

\smallskip Using the above techniques, we are able to have an
approximate, but closed form solution of {\bf SUB-OPT-S}, which allows us to compute the
best $r^*$ that maximizes the detection probability under such an
approximation, as $r^*$ is given in the next section.

% Due to this reason, we analytically study and
% derive the ``optimal'' value of $r^*= \left \lfloor 1+\frac{(1-p)\log
%   K}{2e\log(d-1)}\right\rfloor$ that maximizes the detection
% probability under our framework, which is thus used in our
% algorithm. By selecting this proper $r$, we avoid the issue that needs
% to consider all possible $r$ to maximizes this probability in
% \eqref{eqn:OPT}.

% \note{rationale} \note{how do we simply the complex parts...}

% \smallskip
% \noindent{\bf \em (a) Hop-distance based $C_r$ selection:}
% \note{we solve (a), hop-distance, thus, we can compute... hop-based why
%   easy to compute......high-level..., as will be presented in Theorem 1.},

\subsection{Detection Performance}

We now provide analytical results on the detection performance of {\bf
  SB-Q}$(r).$ We first start by presenting the lower bound
of the detection probability for a given repetition count $r$ in
Proposition~\ref{prop:simple_prob}.

\begin{proposition}
  \label{prop:simple_prob}
  For $d$-regular trees ($d\geq 3$), a snapshot $G_N,$ a given budget
  $K,$ our estimator $\hat{v}(G_N,r)$ from {\bf SB-Q}$(r)$ has the
  following lower-bound of the detection probability:
\begin{multline}
  \label{eqn:detect}
\lim_{N \to \infty}  \bprob{\hat{v}(G_N,r)=v_1}\geq \cr
  \alpha(p,r) \cdot \Bigg(1-c\cdot
  \exp\Bigg( \frac{-h_d
  (K,r)}{2}\Bigg)\Bigg),
\end{multline}
where $\alpha(p,r)=1-\frac{2-p^2}{2}e^{-(p-1/2)^2 \log r}$ and $c = 7(d+1)/d$ and
$$h_d (K,r):=\frac{\log\left(\frac{K}{r}\right)}{\log(d-1)}\log\left(
\frac{\log\left(\frac{K}{r}\right)}{\log(d-1)}\right).$$
\end{proposition}

The proof is presented in Section~\ref{sec:proof}. The second term of RHS of \eqref{eqn:detect} is the probability that the
source is in the candidate set for given $K$ and $r$. Hence, one can see
that for a fixed $K$, large $r$ leads to the decreasing detection
probability due to the smaller candidate set.  However, increasing $r$
positively affects the first term of RHS of \eqref{eqn:detect}, so that
there is a trade off in selecting a proper $r$.

% As a main result, we state the following theorem for the analytic
% performance of Algorithm~\ref{alg:noninteractive}. We first obtain the
% lower bound of detection probability for a given $r$, then by using
% this, we find the optimal $r^*$ and derive that for a given
% $\delta>0$, how many query budget is required to achieve the detection
% probability at least $1-\delta$ as follows.

Now, Theorem~\ref{thm:noninteractive} quantifies the amount of querying
budget that is sufficient to obtain arbitrary detection probability by
choosing the optimal $r^*$ in the sense of the lower bound in
\eqref{eqn:detect}. We provide the proof in Section~\ref{sec:proof}.

% Then, in
% Theorem~\ref{thm:noninteractive}, we present the lower bound on the
% budget for any given detection probability $0< 1-\delta<1.$

% From this result, we
% obtain the best $r^*$ which maximizes the probability in
% \eqref{eqn:detect} and we have the following theorem.
\smallskip
\begin{theorem}
\label{thm:noninteractive}
Using {\bf SB-Q}$(r^*),$ where
$$r^*= \left \lfloor 1+\frac{3(1-p)^2 \log K}{2e\log(d-1)}\right\rfloor,$$
for any given $0< \delta <1,$ the detection
probability under $d$-regular
tree is at least $1-\delta,$ if
\begin{align}
\label{eqn:elower}
     K & \geq \frac{5(d-1)/(d-2) (2/\delta)}{(p-1/2)^2\log (\log (2/\delta))}.
    \end{align}
\end{theorem}

\smallskip This results indicates that if untruthfulness probability is
such that $p=1/2+\varepsilon$ for an arbitrary small number $\varepsilon$,
then we need $1/\varepsilon^2$ times more budget of querying to
satisfy the same target probability. To
illustrate, consider $p=0.7$ and $d=3,$ where we need $K\geq 6156$ to
achieve at least 95\% accuracy of detection.

%\vspace{-0.2cm}

%%% Local Variables:
%%% mode: latex
%%% TeX-master: "main"
%%% End:

\section{Interactive Querying with Direction}
\label{sec:interactive}

In this section, we study the case of interactive querying with
direction. Recall that a source finding algorithm for interactive
queries consists of the following steps: We first need to appropriately
choose the repetition count $r$ and the initial node $v_I$ to ask the
identity question.  If she is not the source then ask the direction
questions $r$ times.  From the answers of querying, the querier chooses
a next node to perform the same procedure. % Hence, the next queried node
% is determined by the previous queried node.

\subsection{Ideal Algorithm and Challenges}
\label{sec:interactive_mle}
Two key components for high detection probability are the choice of the
repetition count and a smart policy which selects the next queriee based
on the answer sample for the direction questions.

Let $\mathcal{P}(v_I)$ be a set of all policies, each of which provides
a rule of choosing a next queriee at each querying step, when the
initial queriee is $v_I.$ Once a policy $P \in \mathcal{P}(v_I)$ and $r$
are chosen, the estimated node $\hat{v}$ is determined, \ie, the node who
reveals itself as the rumor source for the identity question, or the
last queriee, otherwise.  Then, it is natural to choose $r$ and $P$ so
as to solve the following optimization:

\vspace{-0.2cm}
\separator
\vspace{-0.2cm}
\begin{align}
\label{eqn:OPT1}
% \displaystyle
\text{\bf OPT-I:} &  \quad \max_{1\leq r \leq K} \max_{v_I \in
                    V_N}\max_{ P \in \mathcal{P}(v_I)}
\bprob{\hat{v}=v_1|v_I}.
\end{align}
% where the most inner max corresponds to the next node selection
% policy $\mathcal{P}(v_I)$ for a given initial node $v_I$.
\vspace{-0.2cm}
\separator

% \smallskip
% \noindent{\bf \em Challenges.}

We now explain the technical challenges in solving {\bf OPT-I}.  We
first introduce some notations for expositional convenience as well as
our analytical results later.  Let $w_i$ be the $i$-th queriee for
$1\leq i \leq K/r$ \footnote{In out model, $i$ can be strictly less than
  $K/r$ if there exists $i$ such that $i$-th queriee answers ``yes'' for
  the identity question.%  but just to understand t
% there is the source within $K/r$ node selection for
%   querying. However, to understand the optimization problem, we consider
%   that the querier also ask the direction questions to the source so
%   that the all budget $K$ is used.
}, and let $Z_r := Z_r(P) = (z_1, z_2, \ldots, z_{K/r})$ be the sequence
of answers for the identity questions to each queriee for a given
policy $P,$ where $z_i \in \{\text{no},\text{yes}\}$ for the queriee $w_i$. We also let
$D_{r}^{i} := D_{r}^i(q)=[y_1, y_2, \ldots, y_{d}]$ be the answer vector
for the queriee $i$, where $0\leq y_j \leq r$ that represents the number
of ``designations'' to $j$-th neighbor ($1 \leq j \leq d$) as $w_i$'s
parent. % answers of the $j$-th neighbor node by the queried
% node.

As in the simple batch querying, it is important to obtain an analytical
form of the solution of the following problem, to choose the right $r$:
for a fixed $ 1\le r \le K$:
\begin{align}\label{eqn:subOPT1}
% \displaystyle
\text{\bf SUB-OPT-I:} &  \quad \max_{v_I \in V_N}\left(\max_{P \in\mathcal{P}(v_I)}
\bprob{\hat{v}=v_1|v_I}\right).
\end{align}

To solve {\bf SUB-OPT-I}, consider the probability
$\prob{\hat{v}=v_1|v_I}$ in \eqref{eqn:subOPT1} for a given $v_I$.
First, it is pretty challenging to find an optimal policy $P$, because
$P$'s action at each $i$-th queriee can be considered as a
mapping ${\cal F}_i$ that uses the entire history of the queriees and their answers:
% this is difficult to analyze, because it is not easy to obtain
% the optimal policy $\mathcal{P}(v_I).$ To understand this, Then, the
% next queriee selection policy is a function:
\begin{align}\label{eqn:policy}
\mathcal{F}_i : \{D_{r}^{1},D_{r}^{2},\ldots,D_{r}^{i-1}; w_1, \ldots, w_{i-1}\}\rightarrow V_N,
\end{align}
for each $i$. As an approximation, it is natural consider the mapping
$\mathcal{F}_i : (D_{r}^{i-1}, w_{i-1}) \rightarrow V_N$, \ie, the next
queriee is determined only by the information at the moment.  Even under
this approximation, it also remains to estimate the true parent node of
the queriee, using her answers. To handle this issue, we may consider
the MLE to estimate the true parent node, \ie,
$\max_{v \in nb(w_i)} \prob{G_N, D_r^{i-1} | v = \text{parent}(w_i)},$
where $nb(w_i)$ is the set of the neighbors of $w_i.$
However, this is also not easy to analyze, because for some $v$ the
probability that it is a true parent requires to compute the probability
that the true source is located in $v$'s subtree which does not contain
$w_i.$

% the randomness of the snapshot which is related the sum of rumor
% centrality \footnote{Refer our technical report \cite{Jae16} to see the details.}.

Thus, we propose a heuristic algorithm that is designed to produce an
approximate solution of {\bf OPT-I}. The key of our approximate
algorithm is to choose the policy that allows us to analytically compute
the detection probability for a given $r$ so as to compute $r$ easily,
yet its performance is close to that of {\bf OPT-I}.

%Let $pa(v)$ be the true parent of a node $v \in V_N$
%and let $nb(v)$ be the set of all neighbors of $v$.
%Then, we obtain the MLE for the parent of $w_i$ as
%\begin{equation}
%\begin{aligned}\label{eqn:MAP5}
%\hat{w}_{i+1}&=\arg\max_{v \in nb(w_i)}\bprob{G_N,
%D_{r}^{i}|v=pa(w_i)}\\ &\stackrel{(a)}{=}\arg\max_{v \in
%  nb(w_i)}\bprob{G_N,|v=pa(w_i)}\bprob{D_{r}^{i}|v=pa(w_i)},\\
%\end{aligned}
%\end{equation}
%where $(a)$ is from the independence of $G_N$ and $D_{r}^{i}$ for a
%given the event $\{v=pa(w_i)\}.$ In this case, the first term in \eqref{eqn:MAP5} is given
%by $\prob{G_N,|v=pa(v^i)}=\sum_{u \in T_v}R(u,G_N)$ where $T_v$
%is the subtree rooted at node $v$ because if $v$ is the parent of
%$w_i$ then one of node in $T_v$ is the rumor source. The second
%term is Multinomial distribution that $pa(w_i)$ has the probability
%$q$ and $(1-q)/(d-1)$, for other neighbor nodes of $w_i$.

\begin{algorithm}[t!]
 \KwIn{Diffusion snapshot $G_N$, querying budget $K$, degree $d$, and truth
   probabilities $q>1/d$} \KwOut{Estimated rumor source $\hat{v}$}
 Calculate the rumor centrality $R(v, G_N)$ for all $v \in G_N$ as in
\cite{shah2010} and let $ s \leftarrow \arg\max_{v \in V_N} R(v, G_N)$\;
\While{}{ \eIf{$s = v_1$} {
     $K \leftarrow K-1$\;
     Break \;} {\If{$K\geq r$} {Count the number of
       ``designations'' (\ie, She has spread the rumor to me) for the
       direction question among $s$'s neighbors, and choose the largest
       counted node with a random tie breaking\;
       Set such chosen node by $s$\;
       $K \leftarrow K-r$\; } } { } } Return
 $\hat{v}=s$\;
 % \caption{RCP (RC-based Parent tracking) Algorithm \note{change the
 %     format following Algorithm 1.}}
 \caption{{\bf ID-Q}$(r)$: $r$: Repetition Count}
\label{alg:interactive}
\end{algorithm}

\subsection{Algorithm based on Majority Rule}

% We now propose an algorithm that overcomes the afore-mentioned
% complexity.
The key idea our algorithm is that we simply apply a {\em majority-based
  rule}, as will be clarified soon. We first formally describe our
algorithm, called {\bf ID-Q}$(r)$ and explain how it operates and its
rationale. Again, {\bf ID-Q}$(r)$ is parameterized by the repetition
count $r$: we first calculate the rumor
centrality of all nodes in $G_N$ (Line 1), where $s$ is set to be the
rumor center. Then, we query the identity question to the rumor center
$s$ (Lines 3-5).  If she is not the source then ask the direction
questions $r$ times, and count the number of ``designations'' for its
neighbor (Lines 7-8). Then, we choose the largest counted node with a
random tie breaking and repeat the same procedure. The
algorithm stops when there is a node which reveals itself as the rumor
source within $K$ queries, otherwise, it outputs the last queried node
as the estimator.

In selecting a parent node of the target queriee, instead of the exact
calculation of MLE, a simple majority voting is used by selecting the
node with the highest number of designations, motivated by the fact that
when $q>1/d$, such designation sample can provide a good clue of who is
the true parent. Using this idea, we are able to have an
approximate, but closed form solution of {\bf SUB-OPT-I}, which allows us to compute the
best $r^*$ that maximizes the detection probability under such an
approximation, as will be given in the next section.

\subsection{Detection Performance}
We now provide analytical results on the detection performance of {\bf
  ID-Q}$(r)$. We first start by presenting the lower bound of the detection
  probability for a given repetition count $r$ in Proposition~\ref{prop:inter_prob}

\smallskip
\begin{proposition}
  \label{prop:inter_prob}
  For $d$-regular trees ($d\geq 3$), a snapshot $G_N,$ a given budget
  $K,$ our estimator $\hat{v}(G_N,r)$ from {\bf ID-Q}$(r)$ has the
  detection probability lower-bounded by:
\begin{multline}
  \label{eqn:detect2}
\lim_{N \to \infty}  \bprob{\hat{v}(G_N,r)=v_1} \ge \cr
 1-c \cdot \exp\left [-2 (g_d (r,q))^3\left(\frac{K}{r+1}\right)\log \left(\frac{K}{r+1}\right) \right],
\end{multline}
where $g_d (r,q):= 1- e^{-\frac{r(d-1)(q-1/d)^2}{3d(1-q)}}$ and
$c=(8d+1)/d$.
\end{proposition}
\smallskip

The proof is presented in Section~\ref{sec:proof}. The term $g_d (r,q)$ in
\eqref{eqn:detect2} is the probability that the
queriee reveals the true parent for given $r$ and $q$.
% Hence, one can see
% that for a fixed $K$, large $r$ leads to the increasing this
% probability due to the improvement for the quality of the direction answer.
% However, increasing $r$
% negatively affects the term $K/(r+1)$ in \eqref{eqn:detect2}, so that
% there is a trade off in selecting a proper $r$.
Next, Theorem~\ref{thm:interactive} quantifies the amount of querying
budget that is sufficient to obtain arbitrary detection probability by
choosing the optimal $r^*$ in the sense of the lower bound in
\eqref{eqn:detect2}.

\smallskip
\begin{theorem}
\label{thm:interactive}
Using {\bf ID-Q}$(r^*),$ where
$$r^*= \left\lfloor1+\frac{2d(1-q)^2 \log\log
   K}{3(d-1)}\right\rfloor,$$
for any given $0< \delta <1,$ the detection
probability under $d$-regular
tree is at least $1-\delta,$ if
\begin{align}
\label{eqn:elower1}
    K\geq \frac{(2d-3)/d(\log(7/\delta))}{(q-1/d)^3\log (\log (7/\delta))}.
    \end{align}
\end{theorem}

\smallskip The proof is presented in Section~\ref{sec:proof}. This
theorem indicates that if queriees are truthful with probability $q=1/d+\varepsilon$,
for an arbitrary small number $\varepsilon$ we need $1/\varepsilon^3$ times more querying budget.
As an example, suppose $q=0.6$ and $d=3,$ where we need
$K\geq 166$ to achieve at least 95\% accuracy of detection.

It is expected that the querying with direction helps and thus requires less
budget than simple batch querying. Our contribution lies in quantifying
this difference: for small $\delta,$ with respect to the scaling of
$1/\delta$, the amount of querying
is asymptotically reduced from $1/\delta$ to $\log (1/\delta),$ which
should be significant especially when high detection quality is necessary.

% This requires much less queries than that of the simple batch querying
% which reduces the order with respect to $1/\delta$ for small $\delta$,
% from $1/\delta$ to $\log (1/\delta).$

% because the
% answer of direction gives the path information of the rumor source from
% the queriee whereas the batch querying need to consider all nodes within
% the fixed hop distance regardless of the path.

%This result implies that if there is untruthful answer of direction
%with probability $q>1/d$ for the interactive query then it requires
%$1/q^3$ times more budget compared to the case without lying to
%guarantee the target detection probability. Furthermore, if $q\simeq
%1/2$ then, it requires about 8 times more budget for $d$-regular
%tree. Furthermore, if $q=0.4$ then it needs $K\geq 51$ to achieve
%the detection probability at least 95\% for $d=3$.
%Therefore, it requires much less queries than that of batch
%querying which reducing the order of $1/\delta$ to $\log (1/\delta)$
%because the answer of direction gives the path information of the
%rumor source from the queried node whereas the batch querying consider
%all nodes regardless of the path.

%%% Local Variables:
%%% mode: latex
%%% TeX-master: "main"
%%% End:

% \section{Estimating Rumor Sources with Querying}

\section{Proofs of Results}
\label{sec:proof}
% In this section, we will provide the proofs of Propositions and Theorems in
% Section ~\ref{sec:noninteractive} and \ref{sec:interactive}.

\subsection{Proof of Proposition~\ref{prop:simple_prob}}

First, for a given $r,$ we introduce the notation $V_l,$ which is
equivalent to $C_r,$ where the hop distance $l$ is given in Line 3 of
{\bf SB-Q}$(r).$ This is for presentational simplicity due to a
complex form of $l.$ Also for notational simplicity, we simply use $
\prob{\hat{v}=v_{1}}$ to refer to $\lim_{N \to \infty }\prob{\hat{v}(G_N,r)=v_{1}}$
in the proof section.
Then, the detection probability is expressed as the product of the three
terms:
    \begin{align}
      \label{eqn:detection}
      \prob{\hat{v} = v_{1}} &=   \prob{v_1 \in  V_{l}}\times  \prob{\hat{v}=v_1|v_1 \in V_{l}} \cr
                               & = \prob{v_1 \in  V_{l}} \times \prob{v_1
                                 \in \hat{V}|v_1 \in V_{l}}   \cr
                               &  \times \prob{v_1 =v_{LRC}|v_1 \in \hat{V}},
    \end{align}
where $\hat{V}$ is the filtered candidate set (Lines 4-5 of {\bf
  SB-Q}$(r)$) and  $v_{LRC}$ is the node in $\hat{V}$ that has the
highest rumor centrality, where $LRC$ means the local rumor center.
We will drive the lower bounds of the first, second, and the third terms
of RHS of \eqref{eqn:detection}.
The first term of RHS of \eqref{eqn:detection} is bounded by
\begin{align}
  \label{eq:first}
  \prob{v_1 \in V_{l}} \geq 1-c\cdot e^{-(l/2)\log l},
\end{align}
where the constant $c = 7(d+1)/d$ from Corollary 2 of \cite{Khim14}.
The second and the third terms are handled by the following two lemmas,
whose proofs are will be provided in our technical report \cite{Jae16}:
 \vspace{0.2cm}
\begin{lemma}
When $p>1/2,$
\begin{eqnarray*}
\prob{v_1 \in \hat{V}|v_1 \in V_{l}}  & \geq &p+(1-p)(1-e^{-(p-1/2)^{2}\log r}).
\end{eqnarray*}
\label{lem:majority}
\end{lemma}

\smallskip
\begin{lemma}
For a $d$-regular tree ($d\geq3$), if $p>1/2$ then
\begin{equation*}
\begin{aligned}
\prob{v_1 =v_{LRC}|v_1 \in \hat{V}}\geq 1-e^{-p^{2}r\log r}.
\end{aligned}
\end{equation*}
\label{lem:noninteractive}
\end{lemma}

After some algebra, we have lower bounds of $\prob{\hat{v}=v_1|v_1 \in V_{l}}$ by $ 1-\frac{2-p^2}{2}e^{-(p-1/2)^2 \log r}.$ Merging this lower-bound with the lower-bound in
\eqref{eq:first} where we plug in
$l=\frac{\log\left(\frac{K(d-2)}{rd}+2\right)}{\log(d-1)},$ the result
follows. This completes the proof.

\subsection{Proof of Theorem~\ref{thm:noninteractive}}

We first note that the event that our algorithm does not detect the
source to derive the detection error probability is the union of the
following two disjoint events: $E_1:=\{d(v_{RC},v_1) > l\}$  and
$E_2:=\{\hat{v}\neq v_{1}|d(v_{RC},v_1) \leq l\}$, where $E_1$ is the event
that the source is not in the candidate set $V_l$ and $E_2$ is the event
that our estimator fails to detect the source conditioned that the
source is in $V_l.$
% consider the detection error probability for
% given $K$ and $p$. The detection error can be occurred as following two
% cases. First, the source is not in $V_l$ and second, the source is in
% $V_l$ but the estimator can not find the source. To obtain the
% probability, let $E_1:=\{d(v_{RC},v_1) \geq l\}$ be the event for the
% first case and let $E_2:=\{\hat{v}\neq v_{1}|d(v_{RC},v_1) < l\}$ be the
% event for the second case then the error probability is given by
% $\mathbb{P}(\hat{v}\neq v_{1})=\mathbb{P}(E_1 \cup
% E_2)\stackrel{(a)}{=}\mathbb{P}(E_1)+\mathbb{P}(E_2),$
% where $(a)$ is due to the disjoint events of $E_1$ and $E_2$.
Then, from Lemmas~\ref{lem:majority}, we get:
\begin{align}
  \label{eq:ppp}
  \prob{\hat{v}\neq v_{1}} \leq 1-\frac{2-p^2}{2}e^{-(p-1/2)^2 \log r}+c\cdot e^{- \frac{l}{2}\log l},
\end{align}
where the constant $c$ is the same as that in \eqref{eq:first}.
Now, we first put
$l=\frac{\log\left(\frac{K(d-2)}{rd}+2\right)}{\log(d-1)}$ into
\eqref{eq:ppp} and obtained the upper-bound of \eqref{eq:ppp}, expressed
as a function of $r,$ for a given $p,$ and the constant $c.$ Then, we take $r^*$ in the theorem
statement which is derived in \cite{Jae16} and put it to the obtained upper-bound which is expressed as a
function of $K,$ as follows:
\begin{align}
  \label{eq:kkk}
 &  \prob{\hat{v} \neq v_{1}} \cr
&  \leq  \ (1-p)^2 e^{-(p-1/2)^{2}\log K\log (\log K)}+cp^{2} e^{- \frac{\log K}{2}\log (\log K)}\cr
&   \leq  \ (1-p)^2 e^{-(p-1/2)^{2}\frac{\log K}{2}\log (\log K)}+c p^{2}e^{-\frac{\log K}{2}\log (\log K)}\cr
&    \leq  \ c_1 e^{-\frac{(p-1/2)^{2}\log K}{2}\log (\log K)},
\end{align}
where $c_1=c+1$. If we set $\delta \geq c_1 e^{-(p-1/2)^{2}\frac{\log K}{2}\log (\log K)},$ we find the value of $K$ such that its assignment to
\eqref{eq:kkk} produces the error probability $\delta,$ and we get the
desired lower-bound of $K$ as in the theorem statement. This completes
the proof.

\subsection{Proof of Proposition~\ref{prop:inter_prob}}

As a first step, we will obtain the upper bound of the error
probability when $r=1$ \ie, only one direction question is used.
After obtaining this, we can easily extend the result for
general $r$ which will be provided in later.
By similar approach in the proof of Theorem~\ref{thm:noninteractive}, we
define two error events such as $E_1:=\{d(v_{RC},v_1) > K/2\}$ and
$E_2:=\{\hat{v}\neq v_{1}|d(v_{RC},v_1) \leq K/2\}$ which are disjoint.
Then, we have $\prob{E_1}\leq c \cdot e^{- (K/4)\log (K/2)}$
since we use additional direction query with identity question.
Next, by conditioning on the distance $d(v_{RC},v_1) =i$,
the probability for the event $E_2$ is given by
\begin{equation*}
\begin{aligned}
\prob{E_2}&=\sum_{i=1}^{K/2}\prob{\hat{v}\neq
  v_{1}|d(v_{RC},v_1)=i}\prob{d(v_{RC},v_1) =i}.
\end{aligned}
\end{equation*}
We first obtain $\prob{\hat{v}\neq
v_{1}|d(v_{RC},v_1)=i}$ when the total budget is $K.$ To do this, let
$X_j$ be the random variable which takes $+1$ if the answer is
correct at $j$-th direction query by the querier and takes $-1$, otherwise.
Then, the error event is occurred when $\sum_{j=1}^{K/2}X_j <
i$ for all $1\leq j \leq K/2$ because the querier can not meet the source
for the case. Hence, for a given $q>1/d$, we have
\begin{align*}
&\prob{\hat{v}\neq v_{1}|d(v_{RC},v_1)=i}\\ &\leq
\sum_{j=0}^{i-1}\binom{K/2}{j}q^{j}(1-q)^{K/2-j}=I_{1-q}(K/2-i,i)\\ &
\stackrel{(a)}{\leq}
\exp\left(-\left(q-\frac{2i}{K}\right)^2 \frac{K}{2}\right),
\end{align*}
where $(a)$ is due to the Hoeffding bound for the regularized
incomplete beta function $I_{1-q}(K/2-i,i)$ when $q>1/d$. From
the result in \cite{Jae16} for the probability $\prob{d(v_{RC},v_1) =i}\leq \left(\frac{d-1}{d}\right)^{i} e^{- i\log i}$, we obtain
the total probability of error is bounded by
$\prob{\hat{v}\neq v_{1}}\leq c_1 e^{-2q^3
  (K/2)\log(K/2)},$
where $c_1=c+1$. Hence, we conclude the
result for $r=1$. Based on this,
we extend for general $r$. First,
consider the total number of direction queries is $r\geq1$ for each queriee and let
$Y^{1}_i(v)$ be the random variable which takes $+1$ for the $i$-th query
when the true parent node is designated by the queriee $v$ with probability $q$ and let
$Y^{j}_i(v)$ be the random variable which takes $+1$ for the $i$-th query
when one of other neighbor nodes $2\leq j\leq d$ is designated by $v$ with probability
$(1-q)/(d-1)$. Define $Z_j (v)
:=\sum_{i=1}^{r}Y^{j}_i (v)$ be the total number of designations by the node
$v$ for the $j$-the neighbor $(1\leq j \leq d)$. Then, we need to find $\prob{Z_1 (v) > Z_j
(v),~ \forall j}$ which is the probability that the true parent is the node with maximum designations
by queriee $v$. This probability is handled by the following lemma whose proof is given in \cite{Jae16}.

\smallskip
\begin{lemma}
If $q>1/d$ then
\begin{align*}
\prob{Z_1 (v) > Z_j
(v),~ \forall j}\geq 1- e^{-\frac{r(d-1)(q-1/d)^2}{3d(1-q)}}.
\end{align*}
\label{lem:major}
\vspace{-0.5cm}
\end{lemma}
\smallskip
From the result, we obtain
\begin{align}\label{eqn:err}
\prob{\hat{v}\neq v_{1}}&\leq ce^{-2(g_d (r,q))^3
  (K/(r+1))\log(K/(r+1))},
\end{align}
where $c=(8d+1)/d$ and $g_d (r,q):= 1- e^{-\frac{r(d-1)(q-1/d)^2}{3d(1-q)}}$. Hence, we complete
the proof of Proposition~\ref{prop:inter_prob}.

\subsection{Proof of Theorem~\ref{thm:interactive}}

By $r^*$ in the theorem
statement which is derived in \cite{Jae16} and put it to the obtained upper-bound in \eqref{eqn:err} then
\begin{align}\label{eqn:choi}
\prob{\hat{v}\neq v_{1}}&\leq ce^{-2\left(1-
  e^{-\frac{r(d-1)(q-1/d)^2}{3d(1-q)}}\right)^3 (K/(r+1))\log(K/(r+1))}\cr
   &\stackrel{(a)}{\leq}
 ce^{-2(q-1/d)^3 K\log K},
\end{align}
where the inequality $(a)$ is due to the fact that $q>1/d$ with $K\log (K/(r^* +1))>\log K$. Let
$\delta\geq ce^{-2(q-1/d)^3 K\log K}$ then, we obtain the value of $K$ which
 produces the error probability $\delta$ in \eqref{eqn:choi} and we obtain the
desired lower-bound of $K$ as in the theorem statement. This completes
the proof.

%%%% Local Variables:
%%%% mode: latex
%%%% TeX-master: "main"
%%%% End:

%\input{lemma}

\section{Simulation Results}
\label{sec:numerical}

In this section, we will provide simulation results of our two proposed
algorithms over three types of graph topologies: (i) regular trees, (ii)
two random graphs, and (ii) a Facebook graph. We propagate a rumor from
a randomly chosen source up to 400 infected nodes, and plot the
detection probability from 200 iterations.
% and $(c)$ Real world network, respectively. In the simulations, we propagate
% the rumor up to 400 infected nodes and obtain the results by 200 iterations.
\begin{figure}[t!]
\subfigure[SB-Q as varying $p$.]{\includegraphics[width=0.52\columnwidth]{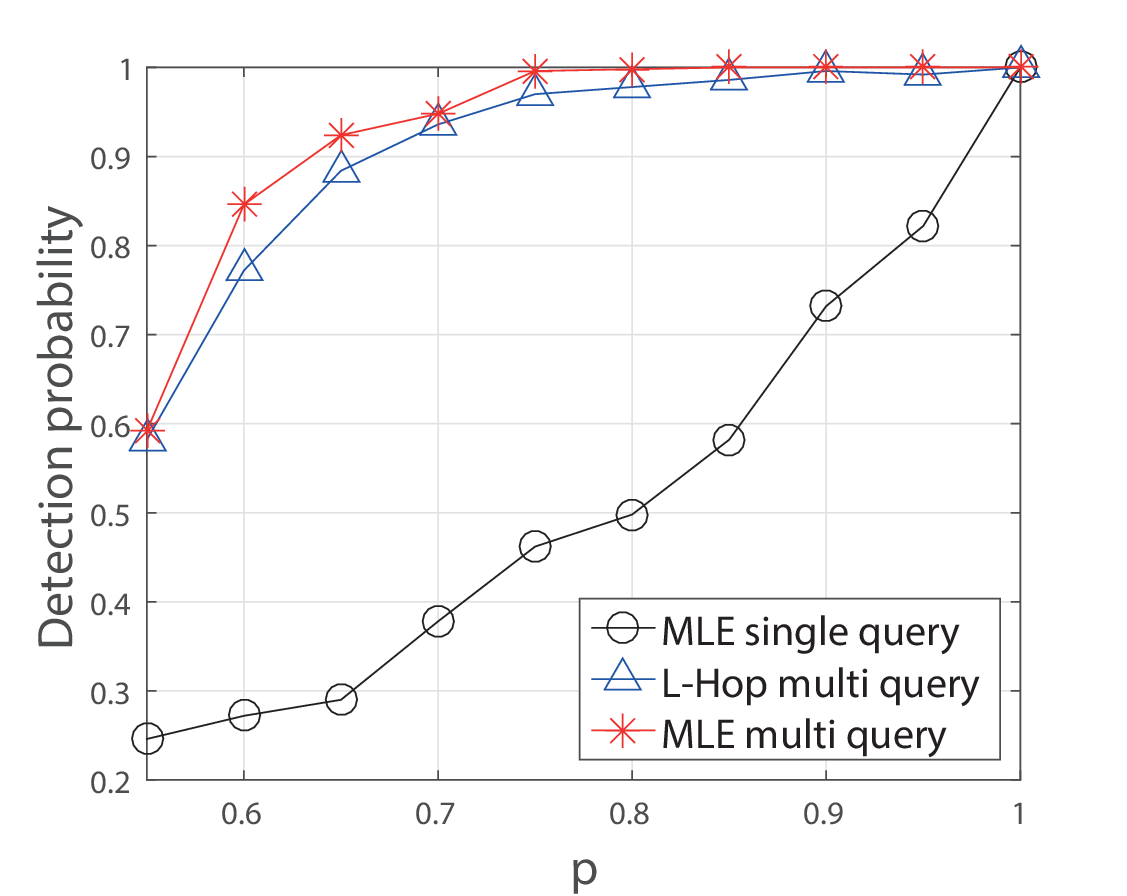}\label{fig:interactive_hop}}
\hspace{-1.5cm}
\subfigure[SB-Q as varying $K$.]{\includegraphics[width=0.52\columnwidth]{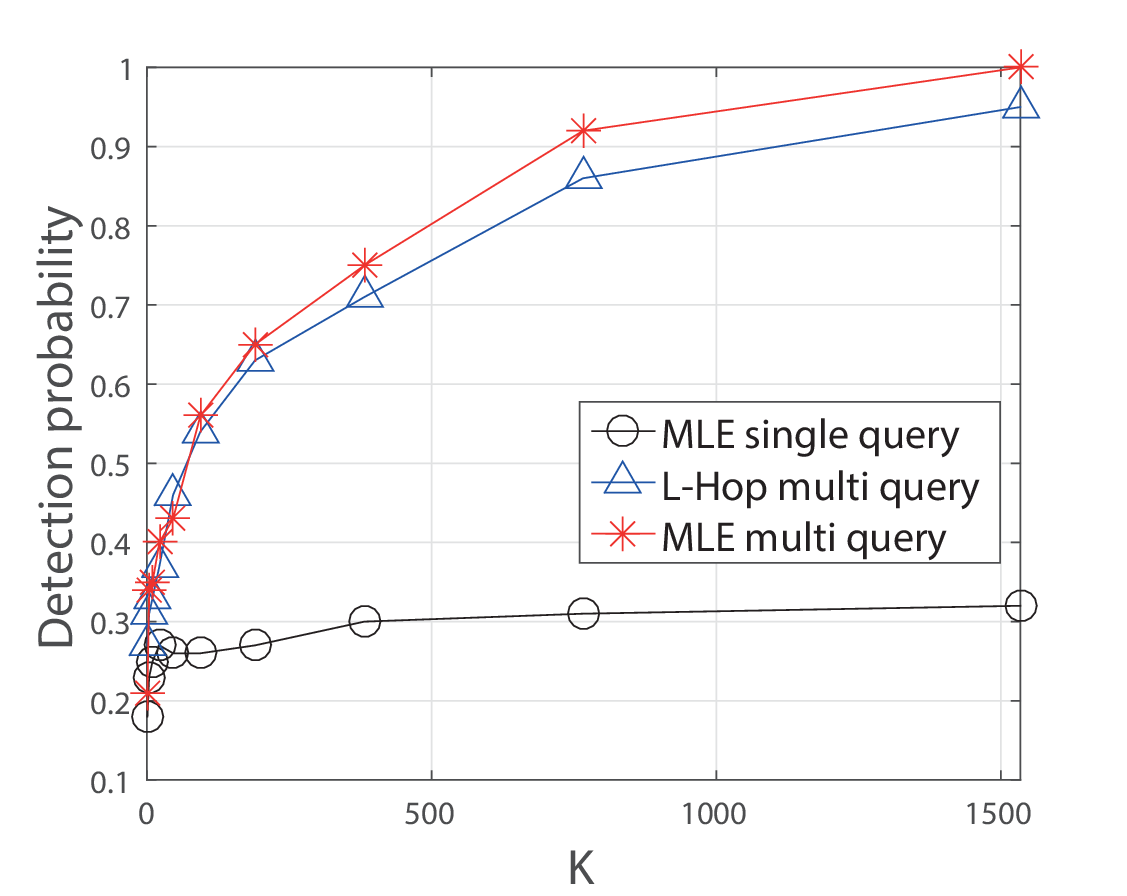}\label{fig:interactive_1hop}}
\subfigure[ID-Q as varying $q$.]{\includegraphics[width=0.52\columnwidth]{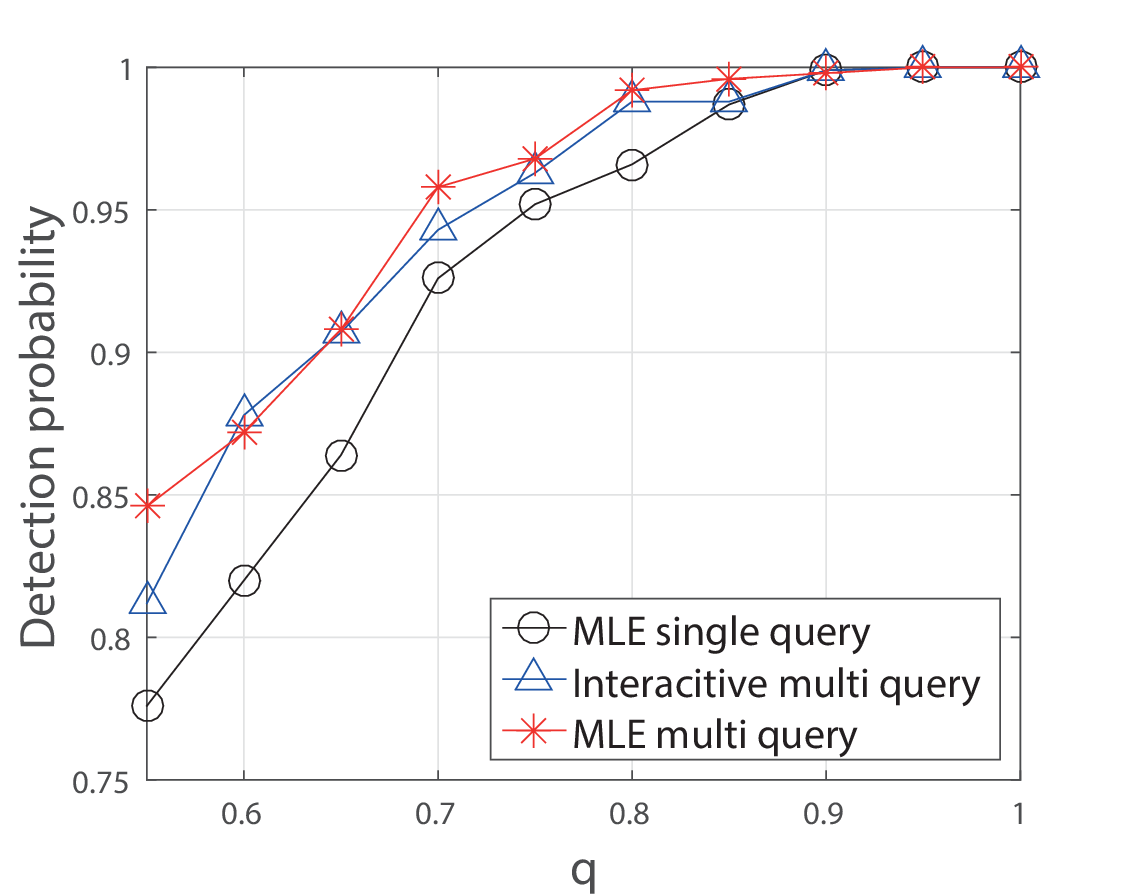}\label{fig:interactive_multihop_a}}
\hspace{-0.6cm}
\subfigure[ID-Q as varying $K$.]{\includegraphics[width=0.52\columnwidth]{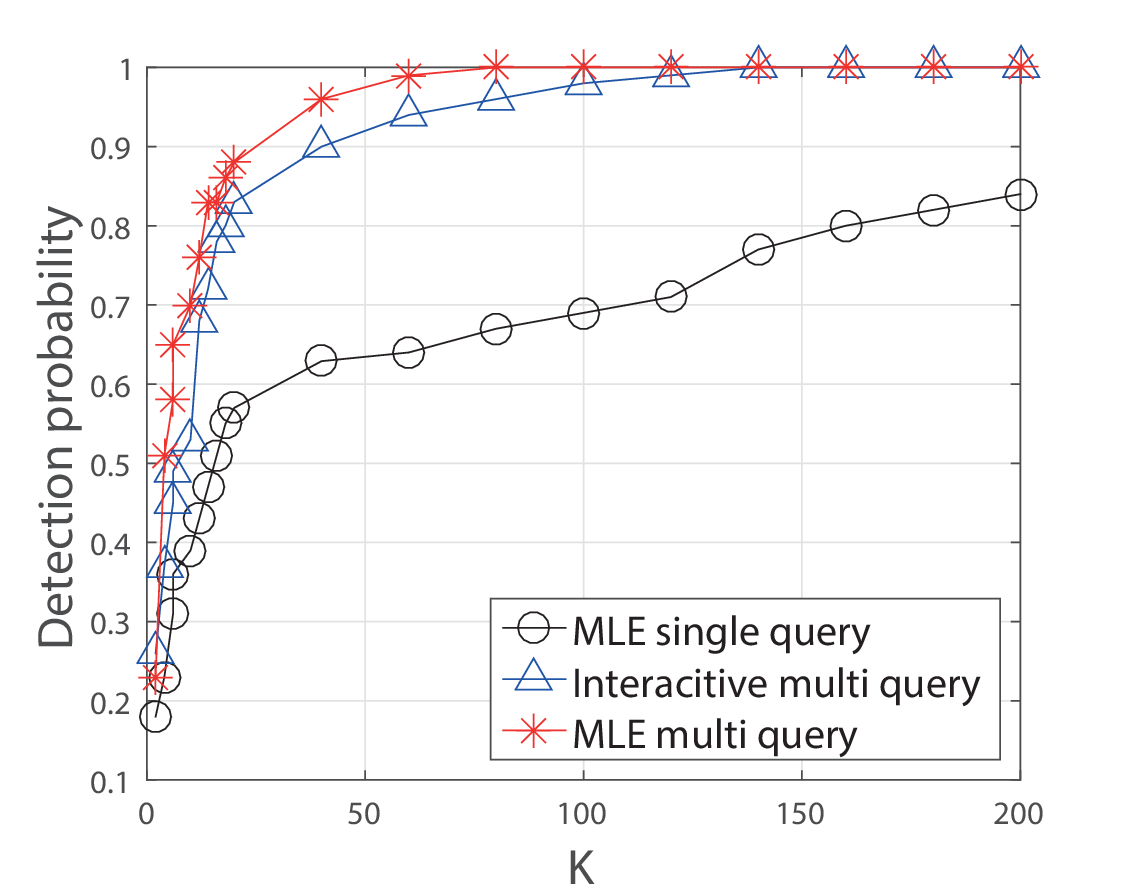}\label{fig:interactive_multihop_b}}
\caption{Detection probabilities for $d$-regular trees ($d=3$) in batch
  querying ((a),(b)) and interactive querying with direction ((c),(d)),
  respectively.}
\label{fig:regular}
\end{figure}

\smallskip
\noindent{\bf \em Regular trees.}
We use $d$-regular tree with $d=3$, where we compare three algorithms
for both simple batch querying (SB-Q) and interactive querying with direction (ID-Q): our algorithms, denoted by
$L$-hop multi query, MLE single query, and MLE multi query.  MLE single
query and MLE multi query are the algorithms that we use $r=1$ and
$r=r^*$ (as described in Theorems~\ref{thm:noninteractive} and
\ref{thm:interactive}), respectively, but for those fixed $r,$ MLE based
estimation algorithms are used as discussed in
Sections~\ref{sec:simple_mle} and \ref{sec:interactive_mle}. Although
MLE multi query is not theoretically optimal, we believe that it is
close to optimal, providing the information on how closely our
algorithms perform compared to optimal ones.
Fig.~\ref{fig:interactive_hop} shows the detection probabilities for
simple batch querying, as the truth probability $p$ varies from 0.55 to
1 when $K=766$ (corresponding to the number of nodes within 8 hop
distance).  As expected, the probability increase as $p$ increases and
we see that if $p=0.7$ then the detection probability is about 40\% for
the MLE single query whereas above 90\% for multi query.  In
Fig.~\ref{fig:interactive_1hop}, we vary the query budget for
$p=0.6$. For multi querying, 1000 queries are enough to achieve the
detection probability is at least 90\% however, it is not beyond 50\%
even for $K=1500$ for the single querying even for MLE, implying that
just selecting a large number of candidate nodes is not enough for
untruthful users, and a certain procedure of learning in presence of
untruthfulness such as multi querying becomes essential.
Figs.~\ref{fig:interactive_multihop_a} and
\ref{fig:interactive_multihop_b} show the similar kind of plots for
interactive querying as $p$ and $K$ vary, respectively.  We observe that
when the detection probability is above 99\% even for MLE single query
if $q>0.9$.  However, for
$q=0.4$ in Fig.~\ref{fig:interactive_multihop_b}, we see that the detection probability is below 90\% when
$K=200$ whereas those of both multi querying schemes are almost one,
showing the power of interactiveness in querying.

% \end{compactenum}
\begin{figure}[t!]
\begin{center}
\subfigure[SB-Q as varying $K$.]{\includegraphics[width=0.517\columnwidth]{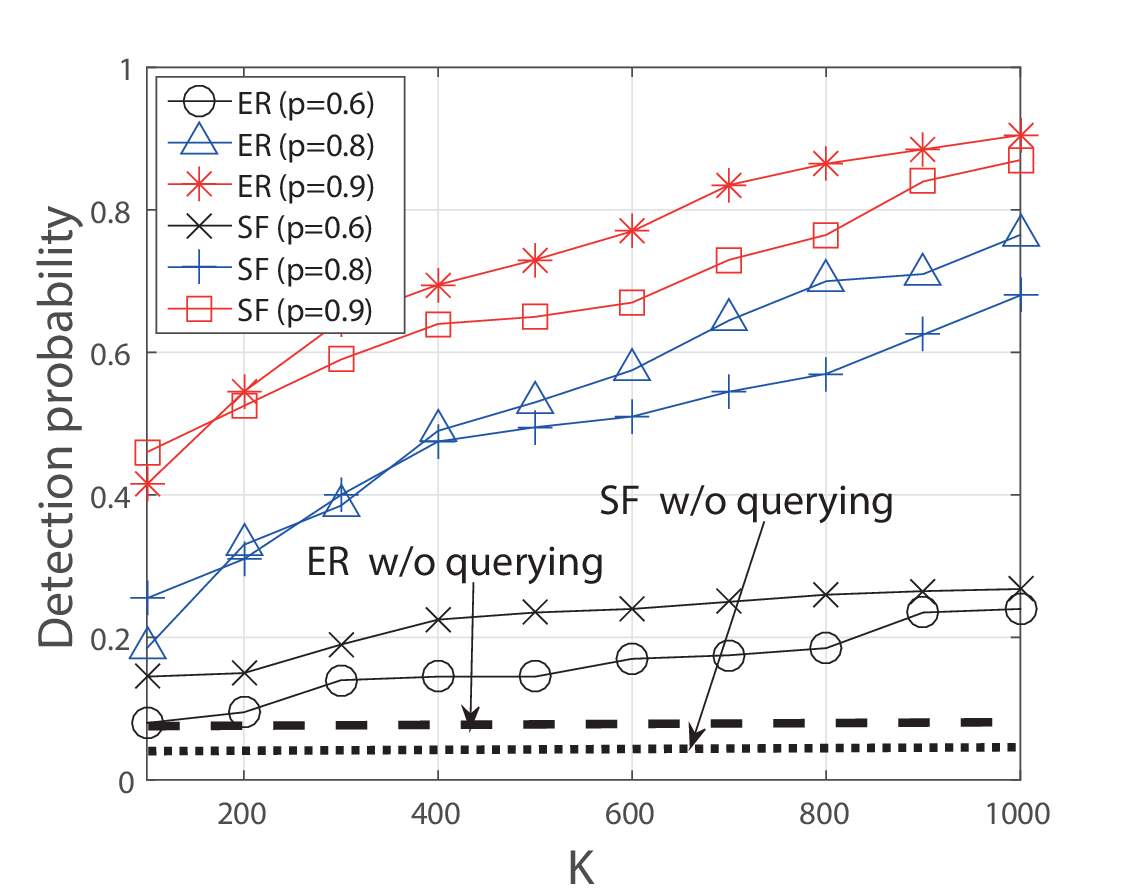}\label{fig:batch_ersf}}
\hspace{-0.55cm}
\subfigure[ID-Q as varying $K$.]{\includegraphics[width=0.517\columnwidth]{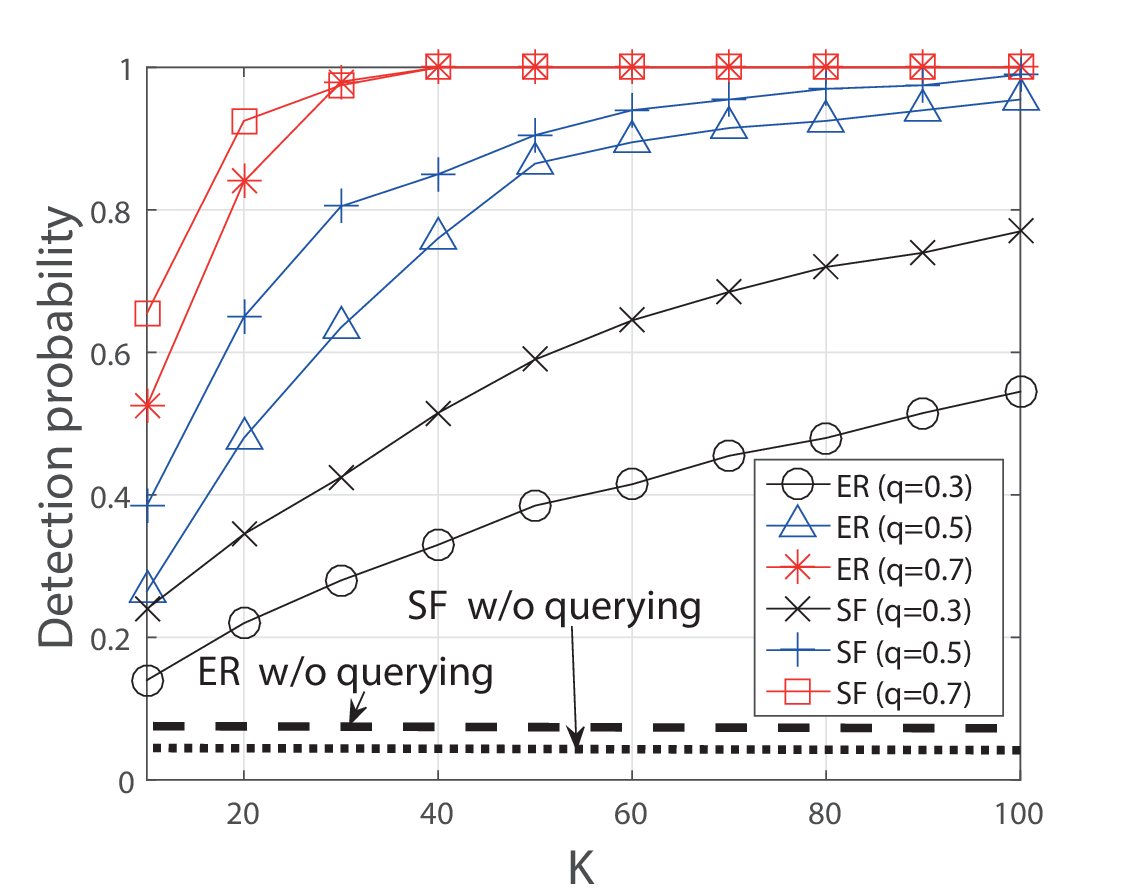}\label{fig:interactive_ersf}}
\end{center}
\caption{Detection probabilities for ER and SF graphs. (Without querying is the detection probability by BFS estimator.)}
\label{fig:random}
\end{figure}

\smallskip
\noindent{\bf \em Random graphs.}
We consider \emph{Erd\"{o}s-R\'{e}nyi} (ER) and scale-free (SF) graphs.
In the ER graph, we choose its parameter so that the average degree by 4
for 2000 nodes. In the SF graph, we choose the parameter so that the
average ratio of edges to nodes by 1.5 for 2000 nodes. It is known that obtaining MLE
is hard for the graphs with cycles, which is $\sharp$P-complete. Due to
this reason, we first construct a diffusion tree from the Breadth-First
Search (BFS) as used in \cite{shah2010}: Let $\sigma_{v}$ be the
infection sequence of the BFS ordering of the nodes in the given graph,
then we estimate the source $v_{\tt bfs}$ that solves the following:
\begin{equation}\label{eqn:BFS}
\begin{split}
  v_{\tt bfs}=\arg \max_{v\in
    G_{N}}\mathbb{P}(\sigma_{v}|v)R(v,T_{b}(v)),
\end{split}
\end{equation}
where $T_{b}(v)$ is a BFS tree rooted at $v$ and the rumor spreads along
it.  Then, by using those selected nodes, we perform our algorithms with
querying.  Figs.~\ref{fig:batch_ersf} and \ref{fig:interactive_ersf}
show the detection probabilities with varying $K$ for batch and
interactive querying, where we observe similar trends to those in
the regular trees. We see that only about 50 questions
need to asked to achieve 99\%
detection probability when $q=0.7$ for the interactive querying scheme.

% detection probabilities for the simple batch querying scheme as
% varying $K$. As we expected, if $p$ is large, then the detection is
% better that of small $p$ for both graphs.
% In Fig.~\ref{fig:interactive_ersf}, we obtain the
% detection probabilities for the interactive querying with direction scheme as
% varying $K$. We see that the direction information is helpful to detect the source
% when $q>1/d$.

% \end{compactenum}

\begin{figure}[t!]
\begin{center}
\subfigure[Facebook Network.]{\includegraphics[width=0.45\columnwidth]{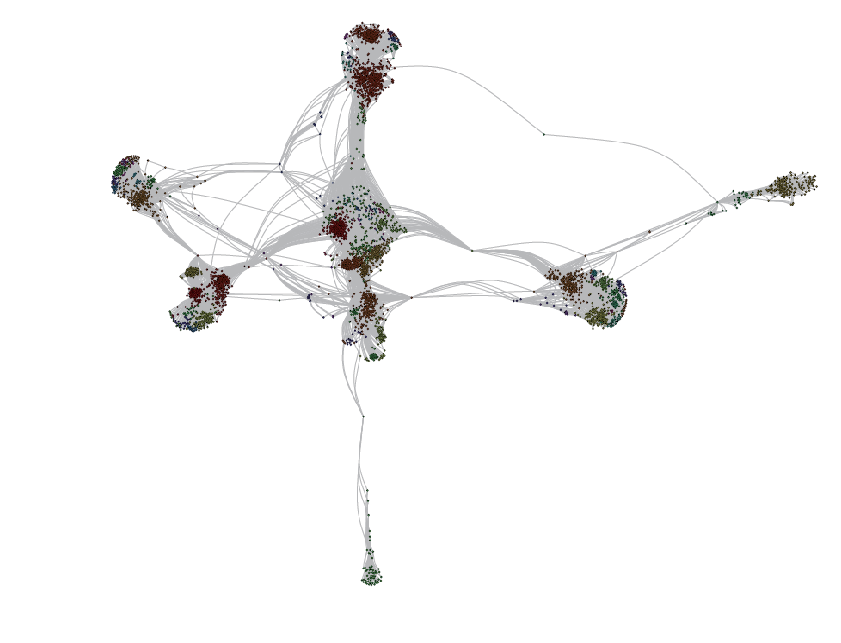}\label{fig:facebook}}
\subfigure[Two querying schemes as varying $K$ in Facebook network.]{\includegraphics[width=0.51\columnwidth]{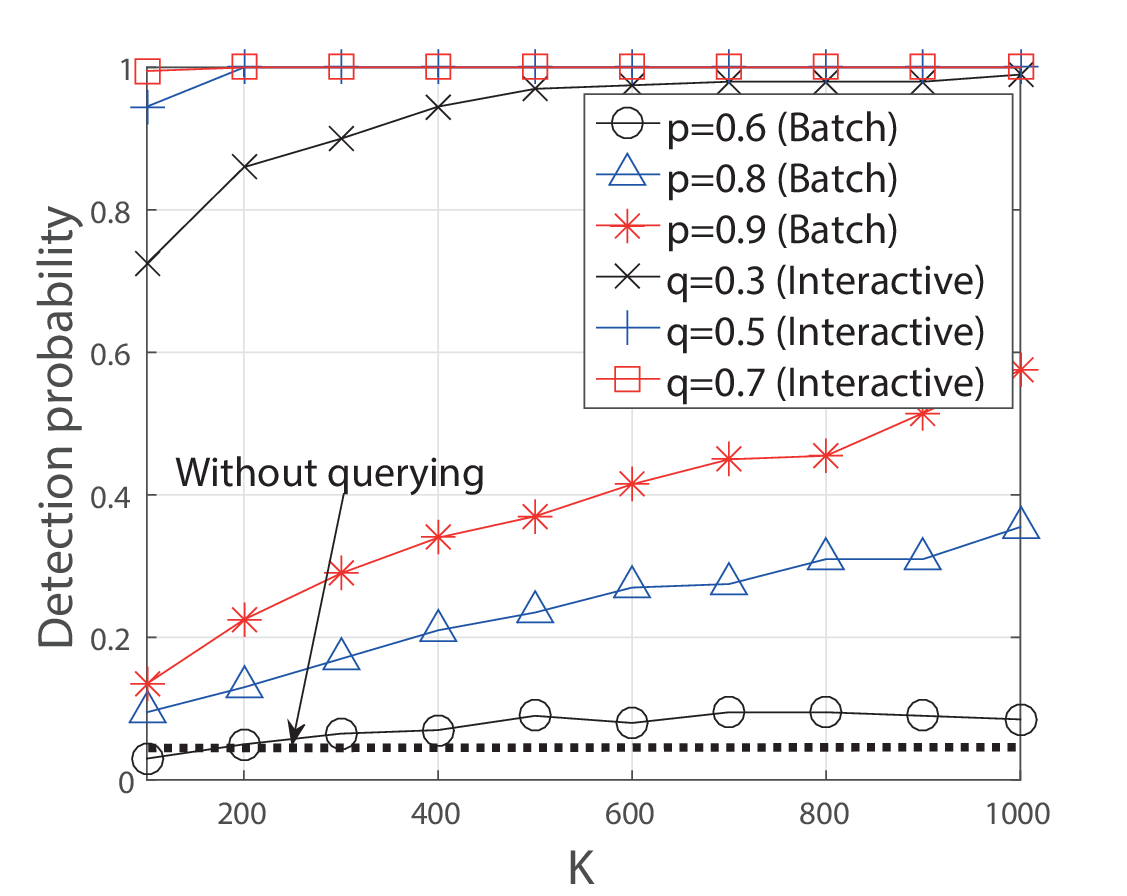}\label{fig:facebook_result}}
\caption{Detection probabilities for Facebook network. (Without querying is the detection probability by BFS estimator.)}
\label{fig:general}
\end{center}
\end{figure}

\smallskip
\noindent{\bf \em Real world graph.}  Finally, we show the results for a
Facebook network as depicted in Fig.~\ref{fig:facebook}.  We use the
Facebook ego network in \cite{NIPS2012} which is an undirected graph
consisting of 4039 nodes and 88234 edges where each edge corresponds to
a social relationship (called FriendList) and the diameter is 8 hops. We
perform the same algorithm used for random graphs based on the BFS
heuristic and show the results in Fig.~\ref{fig:facebook_result}.
The results show that how fast the detection probabilities goes to one as $K$ increases.
For example, the interactive querying requires about 200 queries to achieve almost
one detection probability when $q>0.5$. % because
% the tracking by the direction is efficient due to small diameter of the
% network.
% \vspace{-0.2cm}

%DSBA (Distance Centrality-Based Algorithm)
%\begin{compactitem}[$\circ$]
%  \item {\bf \em DSBA:} dkfdfd
%\end{compactitem}

% \note{TODO: experiment on real-life graph}
% \note{TODO: heuristic algorithm on querying}

%%% Local Variables:
%%% mode: latex
%%% TeX-master: "main"
%%% End:

\section{Conclusion}
\label{sec:conclusion}
In this paper, we have considered the querying framework with
untruthful answers in rumor source detection. We have provided some
theoretical performance guarantees when the underlying network has
regular tree structure.  We obtain how much query budget is required
for two querying types to achieve the target probability
when the truth probabilities are homogeneous in the queriees. 
We perform various simulations based on these algorithms.  As future
works, we will consider the hidden heterogeneous truth probabilities in answers for
both querying scenarios.
% \vspace{-0.3cm}

%%% Local Variables:
%%% mode: latex
%%% TeX-master: "main"
%%% End:

\section*{Appendix}\label{sec:appendix}

\subsection{Proof of Proposition~\ref{pro:subOPT}}
First, note that the {\bf SUB-OPT} is represented by
\begin{align}
  \label{eq:trade1}
&\max_{v \in C_r} \bprob{G_N,A_{r} | v=v_1} \cr
& =  \max_{v \in C_r} \bprob{G_N,A_{r}|v=v_1, v_1 \in C_r} \times \prob{v_1 \in
  C_r}.
\end{align}
We will prove that the RC-based algorithm maximizes for both probabilities
in \eqref{eq:trade1}. First, we consider the second probability.
To see this, suppose $V_{K/r} :=\{v(1),\ldots, v(K/r)\}$ is the
set which contains $K$-largest rumor centrality nodes and let $P(v(i)
=v_1) ~(1\leq i \leq K/r)$ be the source detection probability \ie, the
probability that the $i$-th largest rumor centrality node is the rumor
source. Let $S$ be a set of infected nodes with $|S|=K$ then our
objective is to find
\begin{equation*}
\begin{aligned}
S^* =\arg\max_{S\subset G_N, |S|=K/r}\mathbb{P}(v_1 \in S |G_N).
\end{aligned}
\end{equation*}
Since the probability $P(v_1 \in S |G_N)$ is given by
\begin{align*}
\mathbb{P}(v_1 \in S |G_N)&=\sum_{v \in
  S}\mathbb{P}(v=v_1|G_N)\\ &=\sum_{v \in S}\frac{\mathbb{P}(v=v_1 ,
  G_N)}{\mathbb{P}(G_N)}\\ &=\sum_{v \in S}\frac{P(v=v_1 ,
  G_N)}{\mathbb{P}(v=v_1)}\frac{\mathbb{P}(v=v_1)}{\mathbb{P}(G_N)}\\ &=\sum_{v
  \in
  S}\mathbb{P}(G_N|v=v_1)\frac{\mathbb{P}(v=v_1)}{\mathbb{P}(G_N)},\\
\end{align*}
where $P(G_N)=\sum_{v \in G_N}P(G_N |v)P(v=v_1)$ is independent how
choose the set $S$ and $P(v=v_1)$ is same for all $v \in G_N $. Hence,
we have
\begin{equation*}
\begin{aligned}
\mathbb{P}(v_1 \in S |G_N) &\propto \sum_{v \in
  S}\mathbb{P}(G_N|v=v_1)\propto \sum_{v \in S}R(v, V_N).
\end{aligned}
\end{equation*}
Therefore, the probability $\mathbb{P}(v_1 \in S |G_N)$ is maximum
when $S=V_{K/r}$. Next, we consider the first probability in \eqref{eq:trade1}.
Indeed, this term is also decomposed by
\begin{equation}  \label{eq:trade2}
\begin{aligned}
\bprob{G_N,A_{r}&|v=v_1, v_1 \in C_r}\\
&= \bprob{G_N|v=v_1, v_1 \in C_r}\bprob{A_{r}|v=v_1, v_1 \in C_r}.
\end{aligned}
\end{equation}
Then, one can check that the first probability in \eqref{eq:trade2} is maximized
when the $C_r =V_{K/r}$. Furthermore, the second probability is independent of
set $C_r$ because the querying data is independent to this set. Hence, we obtain
$C_{r}^* = V_{K/r}$ and this completes the proof of
Proposition~\ref{pro:subOPT}.

\subsection{Proof of Lemma \ref{lem:majority}}
Since the case $p=1$ is trivial,
it is enough to show that if $1/2<p<1$ then
    \begin{align}\label{eqn:induct}
     \frac{I_{p}(r-\lfloor r/2\rfloor, \lfloor
       r/2\rfloor+1)-p}{1-p}\geq 1-e^{-(p-1/2)^{2}\log r},
    \end{align}
for any $r\geq 1$. To see this, we use the induction on $r$. First,
for $r=1$, it is holds because the \textbf{LHS} of \eqref{eqn:induct} zero due to
$I_p (1,1)=p$. Clearly, the \textbf{RHS} is also zero. Let
$f_p (r):=(I_{p}(r-\lfloor r/2\rfloor,
\lfloor r/2\rfloor+1)-p)/(1-p)$ and we assume that \eqref{eqn:induct} is holds for $r>1$.
By taking the derivative of $f_p (r+1)$ with respect to $r$, one
can obtain
\begin{equation*}
    \begin{aligned}
    &\frac{\partial(f_p (r+1))}{\partial r}
    \geq\frac{1}{1-p}\left(\frac{\partial I_{p}(r/2, r/2)}{\partial
      r}\right)\stackrel{(a)}{\geq}
    \left(\frac{r+1}{r}\right)^{(p-1/2)^2}-1\\ &\geq
    r^{-(p-1/2)^2}((r+1)^{(p-1/2)^2}-r^{(p-1/2)^2})\\ &\geq
    r^{-(p-1/2)^2}(r+1)^{-(p-1/2)^2}((r+1)^{(p-1/2)^2}-r^{(p-1/2)^2})\\ &=
    \frac{1}{r^{(p-1/2)^2}}-\frac{1}{(r+1)^{(p-1/2)^2}}\\
    &= e^{-(p-1/2)^{2}\log
      r}-e^{-(p-1/2)^{2}\log (r+1)}.\\
    \end{aligned}
  \end{equation*}
where $(a)$ follows from the derivative of the incomplete
function. From the concavity of $I_{p}(r-\lfloor r/2\rfloor, \lfloor
r/2\rfloor+1)$ to $r$, we have
\begin{equation*}
    \begin{aligned}
    f_p (r+1)&\geq f_p (r)+\frac{\partial(f_p (r+1))}{\partial
      r}((r+1)-r)\\ &\geq 1-e^{-(p-1/2)^{2}\log r}\\
      &+(e^{-(p-1/2)^{2}\log
      r}-e^{-(p-1/2)^{2}\log (r+1)})\\ &=1-e^{-(p-1/2)^{2}\log (r+1)},
    \end{aligned}
  \end{equation*}
and this completes the proof of Lemma~\ref{lem:majority}.

%\subsection{Proof of Lemma \ref{lem:noninteractive}}
%
%Let $\mu:=\mathbb{E}[|\hat{V}|]$ be the expected number of candidate nodes in $\hat{V}$.
%Then, we have $\mu=\prob{v_1 \in \hat{V}|v_1 \in V_{l}}+
%(|V_l|-1)(1-\prob{v_1 \in \hat{V}|v_1 \in V_{l}})$. From Lemma~\ref{lem:majority} and
%by using Chernoff bound, we obtain $\mathbb{P}(|\hat{V}|> 1)\leq
%e^{-p^{2}r\log r}$ and this implies
%    \begin{align*}
%      & \prob{v_1 =v_{LRC}|v_1 \in \hat{V}}\cr
%      &\geq\prob{v_1 =v_{LRC}||\hat{V}|=1}\prob{|\hat{V}|=1} \geq 1-
%      e^{-p^{2}r\log r},
%    \end{align*}
%which completes the proof of Lemma~\ref{lem:noninteractive}.
%
%
%
\subsection{Proof of Lemma \ref{lem:major}}
For a given $r\geq 1$, consider each positive constant $\varepsilon_j>0$ for $1\leq j \leq d$. Then, we have
    \begin{align*}
     \prob{Z_1 (v) &> Z_j (v),~ \forall j}=\bprob{\sum_{i=1}^{r}Y^{1}_i (v) >
      \sum_{i=1}^{r}Y^{j}_i (v),~\forall j\neq 1}\\ &\geq
      \bprob{\sum_{i=1}^{r}Y^{1}_i (v) \geq \mu_1
      +\varepsilon_1}\\ &\qquad +\left(1-
      \bprob{\sum_{i=1}^{r}Y^{j}_i (v) \geq \mu_1+\varepsilon_1
      ~\forall j\neq 1}\right)\\ &\stackrel{(a)}{\geq} 1-
      \sum_{j=2}^{d}\prob{\sum_{i=1}^{r}Y^{j}_i (v) \geq
      \mu_j+\varepsilon_j }\\ &\stackrel{(b)}{\geq}
      1-(d-1)e^{-\frac{\varepsilon^2 \mu_j}{ 3}}\geq 1-
      e^{-\frac{\varepsilon^2(d-1)r}{3dq(1-q)}},
    \end{align*}
where $\mu_1=\mathbb{E}[Z_j (v) =\sum_{i=1}^{r}Y^{j}_i (v)]=rq$ and
$\mu_j=\mathbb{E}[Z_j (v) =\sum_{i=1}^{r}Y^{j}_i
  (v)]=r(1-q)/(d-1)$. The inequality $(a)$ comes from the fact that
$\mu_1 \geq \mu_j$ for $2\leq j \leq d$ and the union
bound of probability. From Chernoff-Hoeffding bound of $Y^{j}_i (v)$,
we obtain the inequality $(b)$ by setting $\varepsilon_j =\varepsilon \mu_j$.
If we set $\varepsilon = q^{1/2}(q-1/d)$, we obtain
$\hat{q}\geq 1-e^{-\frac{r(d-1)(q-1/d)^2}{3d(1-q)}},$ which completes the proof
of Lemma~\ref{lem:major}.

%
%
%
%
%%%% Local Variables:
%%%% mode: latex
%%%% TeX-master: "main"
%%%% End:

\balance
{
\bibliographystyle{IEEEtran}
\bibliography{reference}
}
\end{document}